\documentclass[12pt,preprint]{aastex}

\usepackage{epsfig}
\usepackage{graphicx}
\usepackage{color}
\usepackage{enumerate}
\usepackage{ulem}

\newcommand\kms{{\rm km~s^{-1}}}

\def\msun{\,{\rm M_\odot}}

\def\nul{{$-$\tablenotemark{a}}}
\def\nod{$-$\tablenotemark{c}}

\begin{document}

\shorttitle{{Recoiling SMBHs in spin-flip radio galaxies}}
\shortauthors{Liu et al.}
\slugcomment{}

\title{{Recoiling Supermassive Black Holes in Spin-flip Radio Galaxies}}

\author{F.K. Liu\altaffilmark{1}, Dong Wang\altaffilmark{1}, and Xian
  Chen\altaffilmark{2}}

\altaffiltext{1}{Department of Astronomy, Peking University, 100871
  Beijing, China; {\it fkliu@pku.edu.cn}}
\altaffiltext{2}{Kavli Institute for Astronomy and Astrophysics,
  Peking University, 100871 Beijing, China.}

\begin{abstract}
Numerical relativity simulations predict that coalescence of
supermassive black hole (SMBH) binaries not only
leads to a spin flip but also to a recoiling of the merger remnant
SMBHs. In the literature, X-shaped radio sources are popularly
suggested to be candidates for SMBH mergers with spin flip of
jet-ejecting SMBHs. Here we investigate the spectral and spatial
observational signatures of the recoiling SMBHs in radio sources
undergoing black hole spin flip. Our
results show that SMBHs in most spin-flip radio sources have mass ratio
$q\ga 0.3$ with a minimum possible value $q_{\rm min} \simeq 0.05$.
For major mergers, the remnant SMBHs can get a kick velocity as
high as $2100\, {\rm km~s^{-1}}$ in the direction within an angle $\la
40^\circ$ relative to the spin axes of remnant SMBHs, implying that
recoiling quasars are biased to be with high Doppler-shifted broad
emission lines while recoiling radio galaxies are biased to
large apparent spatial off-center displacements. We also calculate the
distribution functions of line-of-sight velocity and
apparent spatial off-center for spin-flip radio sources with different
apparent jet 
reorientation angles. Our results show that the larger the apparent
jet reorientation angle is, the larger the Doppler-shifting
recoiling velocity and apparent spatial off-center displacement will
be. We investigate the
effects of recoiling velocity on the dust torus in spin-flip radio sources
and suggest that recoiling of SMBHs would lead to ``dust poor''
AGNs. Finally, we collect a sample of 19 X-shaped radio objects and
for each object give the probability of detecting the predicted
signatures of recoiling SMBH. 
\end{abstract}

\keywords{black hole physics ---  galaxies: active --- galaxies:
  evolution --- galaxies: jets --- gravitational waves --- quasars:
  general}

\section{Introduction}

In the hierarchical galaxy formation model of the $\Lambda$CDM
cosmology, massive galaxies are produced by successive galaxy
mergers \citep{KH00,spr05}. During the merging of galaxies, the
supermassive black holes (SMBHs) at centers quickly form hard
supermassive black hole binaries (SMBHBs) at a separation of {\it pc}-scale, 
and then may stall for a timescale
longer than Hubble time, if
{ the gravitational potential at galactic nucleus is spherical 
and stellar relaxation is dominated by two-body scattering}
\citep{beg80,yu02}. However, the hardening rate of SMBHBs can be boosted
and SMBHBs become merged within Hubble time
because of triaxial or rotating structures, massive perturbers,
or massive gas disks at galactic centers \citep[][and references
therein]{mm05,col11}. The strong gravitational wave (GW) radiations from
the coalescence of SMBHBs are the main targets of the GW detector Laser
Interferometer Space Antenna (LISA) and the GW detecting program
Pulsar Timing Array (PTA).

Spatially resolved dual or binary active galactic nuclei (AGNs) at
{\it kpc}-scale separation were discovered in the past decade first in
a few merging galaxy systems \citep[e.g.][]{kom03,kom06,gre10}, 
and recently in a large sample of AGNs showing double-peak narrow
emission lines \citep{zho04,XK09,com09,wan09,liu10a,smith10,liu10b,
fu10,shen11,rosario11}. Meanwhile, more compact or hard SMBHBs are 
extremely difficult to observe with present telescopes \citep[see][for
  the only pc-scale SMBHB candidate so far discovered by VLBA]{rod06},
and for sub-pc SMBHBs only indirect observational evidences 
are suggested in a few sub-classes of AGNs with peculiar radio
morphologies \citep{beg80,liu04,liu07}, periodic variabilities
\citep{sil88,val00,liu95,liu97,liu06,liu02}, or special broad-emission-line
features \citep{SF91,gas96,bor09}. Although it is still
unclear how SMBHBs can pass through the final pc-scale
and become merged, some observations of AGNs and
QSOs are suggested to be due to the coalescence of SMBHBs, e.g.
the interruption and restarting of jet activities in
double-double radio galaxies \citep{liu03}, and the rapid jet
reorientation in X-shaped radio sources \citep{mer02,zie05}. For SMBHBs in
quiescent galaxies some observational signatures have been predicted
\citep{che08,che09,che11,KM08b,liu09} but need to be confirmed
observationally.
Among all the observational evidences for the coalescence of SMBHBs in
AGNs, X-shaped radio source is one of the most popular.
X-shaped radio sources are a subclass of double-lobed radio sources,
consisting of a pair of long and diffusive inactive
``wings'' oriented at a large angle relative to the active lobes
\citep{lw84,den02}. The ``spin flip model'' suggests that the 
wings are the fossil plasma of old lobes and
jets are misaligned with the wings because of swift jet reorientation
due to the spin-flip of jet-ejecting SMBHs during the coalescence of
SMBHBs \citep{mer02}. Meanwhile, many other formation
scenarios, which do not require binary black hole coalescence, are
also proposed in the 
literature, including merger of two jet-emitting AGNs \citep{lal07},
hydrodynamic interaction between jet and intra-/inter-galactic medium
\citep{wir82,lw84,wor95,cap02,zie05,SS09,HK10}, and 
precession of jet  axis \citep{eke78,beg80,liu04}.

Numerical simulations show that the coalescence
of two black holes not only leads to a spin-flip
{ black} hole but also to a kick of the post-merger black holes due
to the anisotropic emissions of gravitational waves. Therefore,
it is expected that the coalescence of SMBHBs in spin-flip radio
sources should lead to the recoiling of post-merger SMBHs surrounded
by accretion disk. The recoiling velocity depends on the
angular momentum and mass ratio of the pre-merger SMBHs, and can be
as large as $175\, {\rm km~s^{-1}}$ for Schwarzschild
black holes \citep{gon07},  $3800\, {\rm
km~s^{-1}}$ for extremely spinning equal-mass black hole binaries
with anti-aligned spin angular momenta in the orbital plane
\citep{pre05,her07,bak06,bak08,cam06,cam07a,cam07b}, or
even $10^4 \, {\rm km~s^{-1}}$ in hyperbolic encounters
\citep{hea09}. The gravitational recoiling SMBHs in AGNs may lead to
signatures such as kinematic and spatial offsets
\citep{bon07,kom08,shi09,civ10,rob10,tsa11,era11,MQ04,mag05,
loe07,bat10,jon10,BL08,gue09,civ10,gue11,ble11}, along with some 
other electromagnetic signatures {\citep[see][for a review]{kom10}}.  

In the following, we refer to the radio sources with spin-flip signatures, 
such as those in X-shaped radio sources, as ``spin-flip radio sources''.
In this paper, we adopt the spin-flip model and investigate the
spectral and spatial signatures of the recoiling 
SMBHs in such objects, using the empirical formula obtained in
numerical relativity to calculate the spin angular
momentum and recoiling velocity of the post-merger black hole. We
neglect the slow precession of the spin axes of SMBHs
during the in-spiraling of SMBHBs, We assume that the old and active
jets, respectively, form along the spin axes of the primary and
post-merger SMBHs and any jet-ejecting SMBH has spin parameter $a=c
S / G M^2 \geq 0.9$, where $S$ is the black hole spin angular
momentum, $M$ is the black hole mass, $G$ is the gravitational
constant, and $c$ is the speed of light. For the secondary SMBH, no
constraint on the spin parameter $a_2$ can be given. We assume that
$a_2$ can be any value between 0 and 1 of flat probability distribution.
Our results show that the SMBHBs producing detectable spin-flip angles
should  
form in major mergers with mass ratio $\ga 0.2 - 0.3$ and that the
recoiling SMBHs preferentially move in a direction with an angle $\la
40^\circ$ relative to jet orientation. We calculate the Doppler
kinematic shift of broad emission lines, the AGN
off-center displacement in host galaxy, and the effects of recoiling
SMBHs on dust torus in spin-flip radio galaxies and quasars as a function
of apparent jet reorientation angle and jet size. We finally collect a
sample of 19 X-shaped radio objects and calculate the detection
probabilities for the proposed observational signatures.

The paper is organized as follows. In Section~\ref{sec:method}, we
introduce the empirical formula for the spin angular momenta and the
recoiling velocity of post-merger SMBHs obtained in numerical relativity.
Section~\ref{sec:results} describes our numerical simulations and 
{ general results for spin-flip radio sources}.
We collect a sample of 19 X-shaped radio objects and
calculated the observational signatures of recoiling SMBHs in
Section~\ref{sec:sample}. Our discussions and conclusions are given in
Section~\ref{sec:dc}. Throughout the paper, we assume a $\Lambda$CDM
cosmology with parameters $H_0 = 73\, {\rm km~s^{-1}~Mpc^{-1}}$,
$\Omega_{\rm \Lambda}  = 0.73$, and $\Omega_{\rm m}  = 0.27$.

\section{Empirical formula for spin angular momenta and recoiling
  velocities of post-merger SMBHs}
\label{sec:method}

It is shown that the final spin and the recoiling velocity of a
post-merger SMBH can be calculated in high precision with the
empirical formula fitting the simulation results of numerical
relativity. Following \citet{rez08}, the final spin angular
momentum ${\bf S}_{\rm fin}$ of a post-merger SMBH can be approximated as
the sum of the two initial spin angular momenta $\bf{S}_{1}$ and
$\bf{S}_{2}$ of the primary and secondary SMBHs, respectively, and a
third vector ${\tilde{\bf \ell}}$,
\begin{equation}
  {\bf{S}}_{\rm fin}={\bf{S}}_1+{\bf{S}}_2+{\bf \tilde{\ell}} ,
\label{equ:spin_fin}
\end{equation}
where $\bf{\tilde{\ell}}$ is the difference between the
orbital angular momentum $\bf{L}$ at large SMBHB separation and the
angular momentum $\bf{J}_{\rm rad}$ radiated away during merger
\begin{equation}
  \bf{\tilde{\ell}} = \bf{L} - \bf{J}_{\rm rad} .
\end{equation}
Equation~(\ref{equ:spin_fin}) can be written as
\begin{equation}
  {\bf a}_{\rm fin}=\frac{1}{(1+q)^2} \left({\bf a}_1+{\bf a}_2 q^2 +
  {\bf {\ell}} q \right)
\label{equ:spin_par}
\end{equation}
\citep{rez08},
where mass ratio $q = M_2/M_1 \leq 1$, spin parameter vector ${\bf
a}_{\rm fin}={\bf S}_{\rm fin}/M^2$, ${\bf \ell }=\tilde{\bf{\ell}}
/(M_1 M_2)$, the primary SMBH spin parameter vector ${\bf a}_{1}={\bf
S}_{1}/{M_{1}}^2$, and the secondary SMBH spin parameter vector ${\bf
a}_{2}={\bf S}_{2}/{M_{2}}^2$. $M$, $M_1$, and $M_2$ are,
respectively, the masses of the final SMBH, the primary SMBH, and the
secondary SMBH. Here and in section~(\ref{sec:method}), we use the
natural units $G=c=1$. In equation~(\ref{equ:spin_par}), the direction
of the angular momentum vector $\bf{{\ell}}$ is nearly parallel to the
orbital angular momentum and its norm is
\begin{eqnarray}
\vert {\bf {\ell}} \vert & = & \frac{s_4}{(1+q^2)^2} \left(\vert {\bf
  a}_{1} \vert^2 + \vert {\bf a}_{2} \vert^2 q^4
+ 2 \vert {\bf a}_{1} \vert^2 \vert {\bf a}_{2} \vert^2 q^2
\cos\alpha\right) +         \nonumber \\
&&
\left(\frac{s_5 \eta + t_0 + 2}{1+q^2}\right) \left(\vert {\bf
  a}_{1} \vert\cos\beta + \vert {\bf a}_{2} \vert q^2
\cos\gamma\right) + \nonumber \\
&&
2 \sqrt{3}+ t_2 \eta + t_3 \eta^2 \, ,
\label{equ:ang_diff}
\end{eqnarray}
where $s_4$, $s_5$, $t_0$, $t_2$, and $t_3$ are the empirical
fitting coefficients given by \citet{rez08}, $\eta$ is the symmetric
mass ratio $\eta \equiv M_1M_2/(M_1+M_2)^2 = q/(1+q)^2$, and the three
projected (cosine) angles $\alpha, \beta$ and $\gamma$ are defined
with the inner products
\begin{equation}
  \cos \alpha \equiv
       {{\bf \hat{a}}_1\cdot{\bf \hat{a}}_2} \,,
\hskip 0.3cm
\cos \beta \equiv {\bf \hat a}_1\cdot{\bf \hat{{\ell}}}\,,
\hskip 0.3cm
\cos \gamma \equiv {\bf \hat{a}}_2\cdot{\bf \hat{{\ell}}}\,.
\label{equ:proj_angle}
\end{equation}
In the definition of $\bf{{\ell}}$, the orbital angular momentum
$\bf{L}$ is the one at large (infinity) separation. In the
calculation, we neglect the slow precession of spin axis around total
angular momentum and use all the empirical formula for a finite
separation $\gg M$ to calculate the spin-flip angle. We expected that
this approximation can give good enough results to describe the jet
reorientations in X-shaped radio sources.

Coalescence of SMBHBs not only leads to the spin-flip but also to a
fast recoiling of the post-merger SMBH. The recoiling velocity of
post-merger SMBH can be empirically given by
\begin{eqnarray}
{\mathbf V} &=& V_{m} \, {\mathbf e}_1 + V_{\perp s} (\cos\xi \,
    {\mathbf e}_1 + \sin\xi \, {\mathbf e}_2) + V_{\parallel s} \,
    {\mathbf e}_3,
\label{equ:recoil_vel}\\
    V_{m}    &=& A \eta^2 \sqrt{1 - 4 \eta} (1 + B \eta),
\nonumber\\
    V_{\perp s}     &=& H \frac{\eta^2}{(1+q)} \left( a_1^{\parallel}
    - q a_2^{\parallel} \right)
\label{equ:recoil_velv}
\end{eqnarray}
\citep{cam07a,gon07},
where ${\mathbf e}_1$ and ${\mathbf e}_3$ are, respectively, the unit
vectors in the directions of the binary separation from the
  primary to the secondary black hole and the orbital
rotating axis just before merger, and the unit vector
${\mathbf e}_2\equiv{\mathbf e}_1\times{\mathbf e}_3$. The index
$\parallel$ and $\perp$ refer to, respectively, the projections of
the vectors parallel and perpendicular to the orbital axis. The
index $m$ indicates the recoil velocity of unequal mass contribution,
and $s$ indicates the contribution due to spin. The fitting parameter
$\xi$ is the angle between the recoiling velocities in the orbital
plane due to the mass asymmetry and spin angular momenta and we use
$\xi = 145^{\circ}$ in the following calculations as in
\citet{lou08}. The fitting coefficients $A$, $B$, and $H$ are taken
from \citet{cam07a}. For the kick velocity parallel to orbital
axis, we use the empirical formula given by \citet{van10}. These formula
are equivalent to one suggested by \citet{lou10} if angular parameters are
suitably interpreted. In order to use these formula for statistical
calculation, we combined the equations (7), (8) and (9) in \citet{van10}
and adopted the interpretation of angular parameters from \citet{lou10}.
Then we obtained the following formula
\begin{eqnarray}
  V_{\parallel s} &=&
  -\frac{K_2\eta^2+K_3\eta^3}{q+1}|a^{\perp}_1-qa^{\perp}_2|
   \cos(\Theta_{\Delta}-\Theta_0)\nonumber\\
  &+&\frac{K_S(q-1)\eta^2}{(q+1)^3}|a^{\perp}_1+q^2a^{\perp}_2|
  \cos(\Theta_{S}-\Theta_1),
\label{equ:recoil_velp}
\end{eqnarray}
where $K_2$, $K_3$, and $K_s$ are fitting parameters taken from \citet{van10}.
In equation~(\ref{equ:recoil_velp}), $a^{\perp}_i$ is the magnitude of
the spin parameter vectors ${\bf a}_{\rm i}$ of the $i$th black hole
projected into the orbital plane, $\Theta_{\Delta}$
($\Theta_{S}$) is the angle between the in-plane component of
$\Delta\equiv \mathbf{S}_2/M_2-\mathbf{S}_1/M_1$ ($\mathbf{S}\equiv
\mathbf{S}_1+\mathbf{S}_2$) and the unit separation vector from
the primary toward the secondary at merger and $\Theta_0$ ($\Theta_1$)
is the corresponding initial value of $\Theta_{\Delta}$ ($\Theta_{S}$)
at the arbitrary separation of the binary with quasi-circular
orbit.

%----------------------
\section{Numerical Simulations and Results}
\label{sec:results}

%---
\subsection{Monte Carlo simulations}
\label{sec:MC}

With the empirical formula given in Section~\ref{sec:method}, we now
investigate the distributions of the SMBH mass ratio, spin-flip angle,
recoiling velocity in spin-flip radio sources under the
spin-flip model. In our calculations, the physical quantities are
specified in the Cartesian coordinate system defined by ${\bf e}_1$,
${\bf e}_2$, and ${\bf e}_3$ and so that the positive direction of the
$z$-axis align with the orientation of ${\bf e}_3$ and the orbital
plane of the pre-merger SMBHB is in the $x$--$y$ plane. To calculate
the final spin vector ${\bf a}_{\rm fin}$ and the recoiling velocity
$\bf{V}$ for a post-merger SMBH, we have to specify the following
physical quantities of the pre-merger SMBHB: $q$,
$\Theta_{\Delta}$,  $\Theta_{S}$, $\Theta_0$, $\Theta_1$, and 
both the magnitude and orientation of ${\bf a}_1$ and ${\bf a}_2$.
We employ the Monte Carlo scheme to generate these quantities
according to the statistical properties motivated by the physical
considerations. The details are as follows.

The mass ratio $q$ of the SMBHB has a uniform distribution in the
range $[0, 1]$, and the spin vectors ${\bf a}_1$ and ${\bf a}_2$ are
randomly oriented {(unless otherwise noted)}. Following \citet{lou10} and for statistical
purpose, in the simulations we define $\Theta_{\Delta}$ and 
$\Theta_{S}$ with respect to ${\bf e}_1$, and take $\Theta_0$ = 0,
$\Theta_1$ has a uniform distribution in the range $[0, 2\pi]$.
{ In the case of an spin-flip radio source,} although both SMBHs 
before merger could be the candidates
for the jet-ejecting black holes, we assume that the low luminosity
diffusive lobes are
produced by the primary SMBH before coalescence and the active jets
form along the final spin axis of the post-merger SMBH. Because we are
interested in radio loud AGNs or quasars with powerful
relativistic jets both before and after the merger, we do simulations
only for uniform distributions of $a_1$ within the range $0.9 \leq a_1
\leq 0.998$. For the spin of the secondary SMBH, we cannot give any
priori constraint on it thus we assume a uniform distributions in the
range of $0\leq a_2 \leq 0.998$. Among the simulations, we adopt only
those corresponding to $0.9 \leq a_{\rm fin} \leq 0.998$ as the final
results for  spin-flip radio sources. Although we set an upper limit
0.998 rather than the extreme spin $a=1$ to all the spin parameters
by physical consideration, our results are not significantly
changed by this small difference in upper limit because of the
regularity of the results and the rarity of SMBH merger at $a=1$.

In each run of Monte Carlo simulations, we first draw a set of
physical quantities according to the distribution functions specified
above, and then calculate ${\bf a}_{\rm fin}$ and $\bf{V}$, using the
equations~(\ref{equ:spin_par})-(\ref{equ:recoil_velp}). To
derive the distribution functions of the physical quantities, we use
$10^7$--$10^9$ sets of the quantities, depending on the request of
resolution. To compare the
numerical results with the real observations, we project the vectors
${\bf a}_{\rm fin}$ and $\bf{V}$ both along and vertical to the line
of sight (LOS) vector $\bf{e}_{\rm s}$, which is assumed to randomly
orient. We calculate the angles $w_1$ and $w_2$, respectively, between
${\bf a}_1$ and $\bf{e_s}$ and between ${\bf a}_{\rm fin}$ and
$\bf{e_s}$, which may be observed directly or observationally
constrained. Because we assume that  jets
form along the spin axes of the primary SMBHs before merger and of the
final SMBHs after merger, the spin flip during merger leads to the
reorientation of jets. We also calculate the projected (apparent) jet
reorientation angle $\delta$, i.e. the angle between the components of
${\bf a}_1$ and ${\bf a}_{\rm fin}$ projected vertically to LOS
\begin{equation}
\cos(\delta) = \frac{({\bf a}_1\times\bf{e_s})\cdot({\bf a}_{\rm
    fin}\times\bf{e_s})}{\vert{\bf a}_1\times\bf{e_s}\vert\vert{\bf
    a}_{\rm fin}\times\bf{e_s}\vert} .
\label{equ:reorient}
\end{equation}

{ To be identified as a spin-flip radio source, both the
active jets forming along the spin axis of the post-merger SMBH
and the relic jets forming along the spin axis of the pre-merger primary SMBH
should be resolvable by radio telescopes. Moreover, the active jets 
should orient at a large angle with respect to the relic jets. 
In the special case of X-shaped radio sources,}
the length of the relics jets (wings) should be at least $80\%$ of the
length of the active ones, and the angle between relic and active jets 
should exceed $15^\circ$ \citep{LP92}. The requirements of length and
symmetry imply that the angle between relic jets and LOS cannot be very
small, and clear identification of diffusive extended wings from the
plasma cocoon of active jets also requires a large relative angle
between the projected wings and active lobes in the plane of the sky.
Therefore, the simulations resulting in $w_1 < 20^\circ$ or
$\delta < 15^{\circ}$ are not included in the later statistical
analysis. Besides, a radio source would be observed as a blazar if the
relativistic jet is nearly along the line of sight and the
relativistic beaming effect becomes dominant. 
{ Resolving active jets and good measurement of their length
and orientation therefore require that $w_2$ cannot be very small.
For this reason,} the simulations resulting in $w_2 < 10^\circ$ are excluded.

After excluding the solutions of the equations which do not
fulfill the above selection criteria, only about $10^5$-$10^6$
simulations are left and will be used in the derivation of the
distribution
functions of the SMBH mass ratio $q$, the spin flip angle $\Delta$,
and recoiling velocity $\bf{V}$ for different apparent jet
reorientation angles $\delta$, and in the investigation of the effects
of recoiling velocity on dust torus. To analyze the
standard deviation of a distribution function, we run in total 5
sets of Monte Carlo simulations.

%%%%%%%%%%%%%
\subsection{Distributions of SMBH mass ratios in Spin-flip
  radio sources}
\label{sec:mratio}

\begin{figure}
\plotone{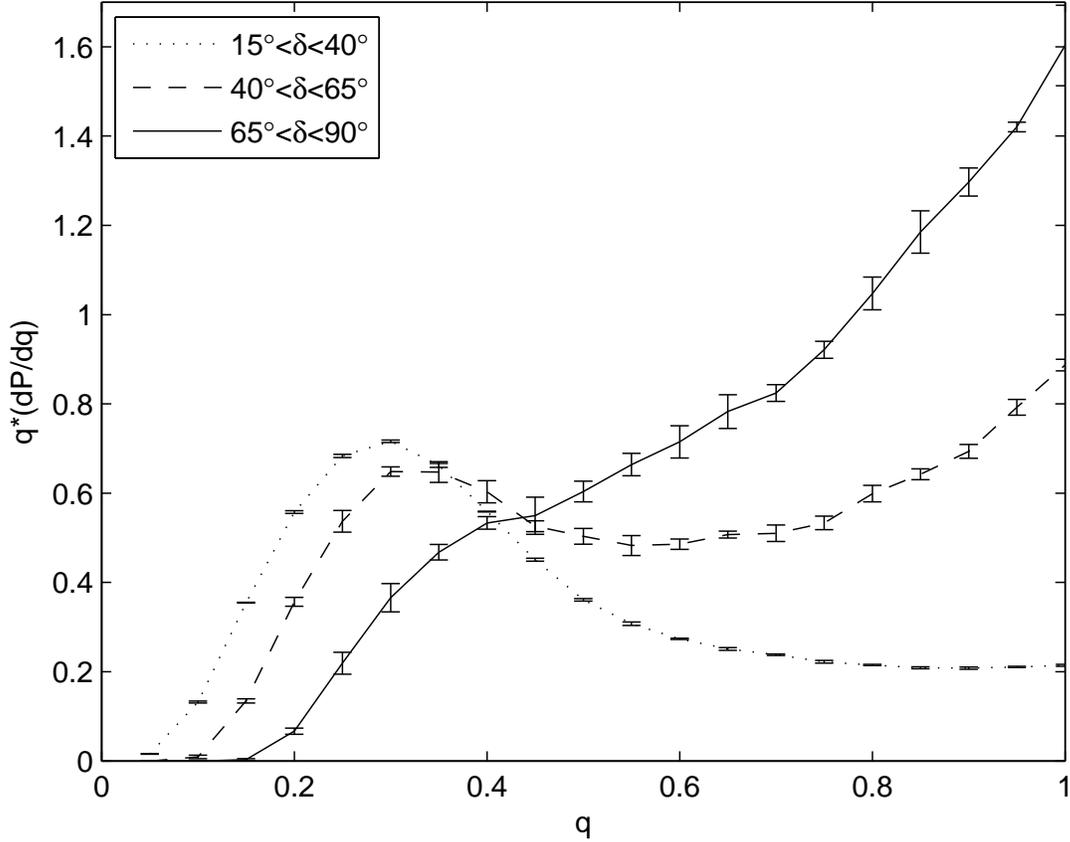}
\caption{Distribution function of SMBHB mass ratio $q$ in spin-flip
  radio sources for different apparent jet reorientation angles
  $\delta$. Apparent jet reorientation angle $\delta$ is the angle
  between the orientations of the active and remnant (wings) jets,
  which is also the projected angle between the spin axes of the
  primary and the post-merger SMBHs. The minimum mass ratios depend on
  $\delta$ and are, respectively, $q_{\rm min}\simeq 0.05$ for
  $15^{\circ} <\delta <40^{\circ}$, $q_{\rm min}\simeq 0.1$ for
  $40^{\circ} <\delta
  <65^{\circ}$, and $q_{\rm min}\simeq 0.15$ for $65^{\circ} <\delta
  <90^{\circ}$. SMBHBs in most spin-flip radio sources have mass ratio
  $q\ga 0.3$ and thus are formed by major mergers.
  Error bar gives $1\sigma$ standard error.}
\label{qdis}
\end{figure}

Our results of the distribution of SMBH mass ratio $q$ in the simulated
{ spin-flip} radio sources are shown in Figure~\ref{qdis}. We
calculate the apparent jet reorientation angle using
Equation~(\ref{equ:reorient}) and divide the simulation results into
three groups, based on the apparent jet reorientation angles
for $15^{\circ}<\delta<40^{\circ}$, $40^{\circ}<\delta<60^{\circ}$,
and $65^{\circ}<\delta<90^{\circ}$. Given the conditions we have,
the mass ratio $q$ in spin-flip radio sources cannot be uniquely
determined, therefore we are able to give only the statistical
probability $q (dP/dq)$ of a certain mass ratio, as the
distribution function in Figure~\ref{qdis}.

Our results imply that to form a detectable apparent jet reorientation
angle with $\delta \ge 15^{\circ}$, the mass ratio of the SMBHB must
be larger than $0.05$, which is smaller than the minimum mass ratio
obtained by \citet{HB03}. This difference of the minimum mass ratio
is due to the projection effects on the spin flip which
can amplify a small jet reorientation angle. If we set
smaller lower limits on $w_1$ and $w_2$, the amplification effect
becomes more significant and we may get an even smaller minimum mass
ratio. Our results also show that for a larger apparent jet reorientation
angle the needed minimum mass ratio becomes larger. To form spin-flip
radio structure with apparent jet reorientation angle $40^{\circ} < \delta <
60^{\circ}$, the needed minimum mass ratio is $q_{\rm min} \simeq
0.1$, while for $65^{\circ} < \delta < 90^{\circ}$, $q_{\rm min}
\simeq 0.15$.

Figure~\ref{qdis} showed that for an spin-flip radio source with
apparent jet
reorientation angle $15^{\circ} <\delta <40^{\circ}$, the SMBHB most
probably has mass ratio $q \sim 0.3$, and large contribution comes
from SMBHBs with
mass ratio $0.15 \la q \la 0.5$ although the contribution from 
SMBHBs with $q\ga 0.5$ cannot be completely neglected.
For spin-flip radio sources with larger apparent jet reorientation
angles, the contribution of the mergers of SMBHBs with larger mass
ratio $q$
become more significant while that of SMBHBs with smaller mass ratio
becomes less important. For spin-flip radio sources with intermediate
apparent jet reorientation angle $40^{\circ} <\delta <60^{\circ}$, the mergers
of SMBHBs with mass ratio $q \ga 0.25$ may make significant
contributions, although the contribution from SMBHB mergers with $q\sim
0.5$ is at local minimum. For $65^{\circ} < \delta <90^{\circ}$,
the jet reorientation in spin-flip radio sources is likely due to the merger of
SMBHBs with mass ratio $q \ga 0.35$, the larger the mass ratio, the
more likely, and the largest contribution is due to the mergers with
$q\ga 0.6$. We notice that for $65^{\circ} < \delta <90^{\circ}$ the
typical probability $q(dP/dq)$ could be larger than unit although the
integrated 
probability is unit. This is due to the reason that the probability
per unit mass ratio varies very rapidly on a length scale much
less than the typical mass ratio or $P/(dP/dq) \ll q$ for $q
\ga 0.8$.

%---
\subsection{Spin flip angles and recoiling velocities of post-merger
  SMBHs}

\begin{figure}
\plottwo{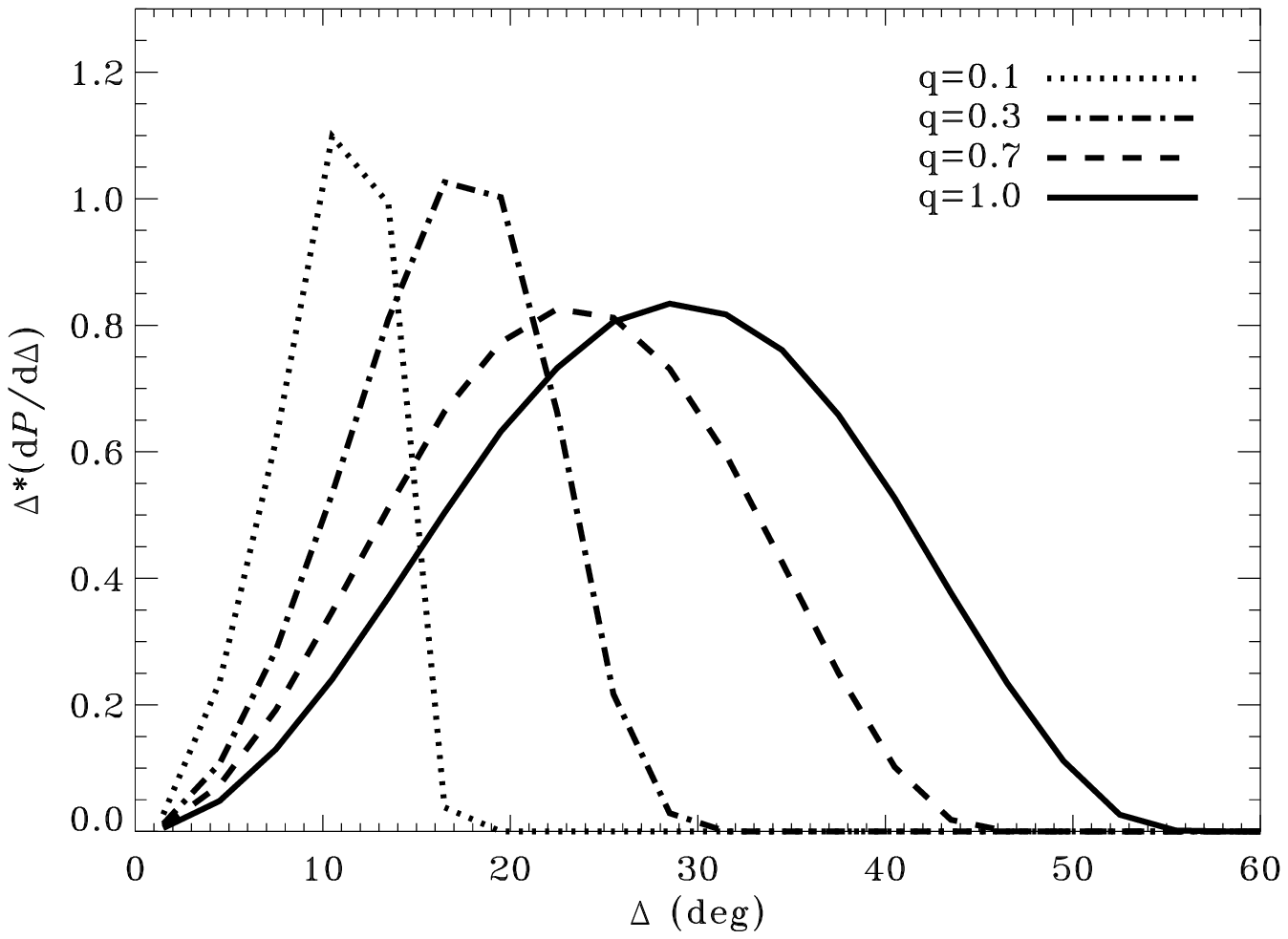}{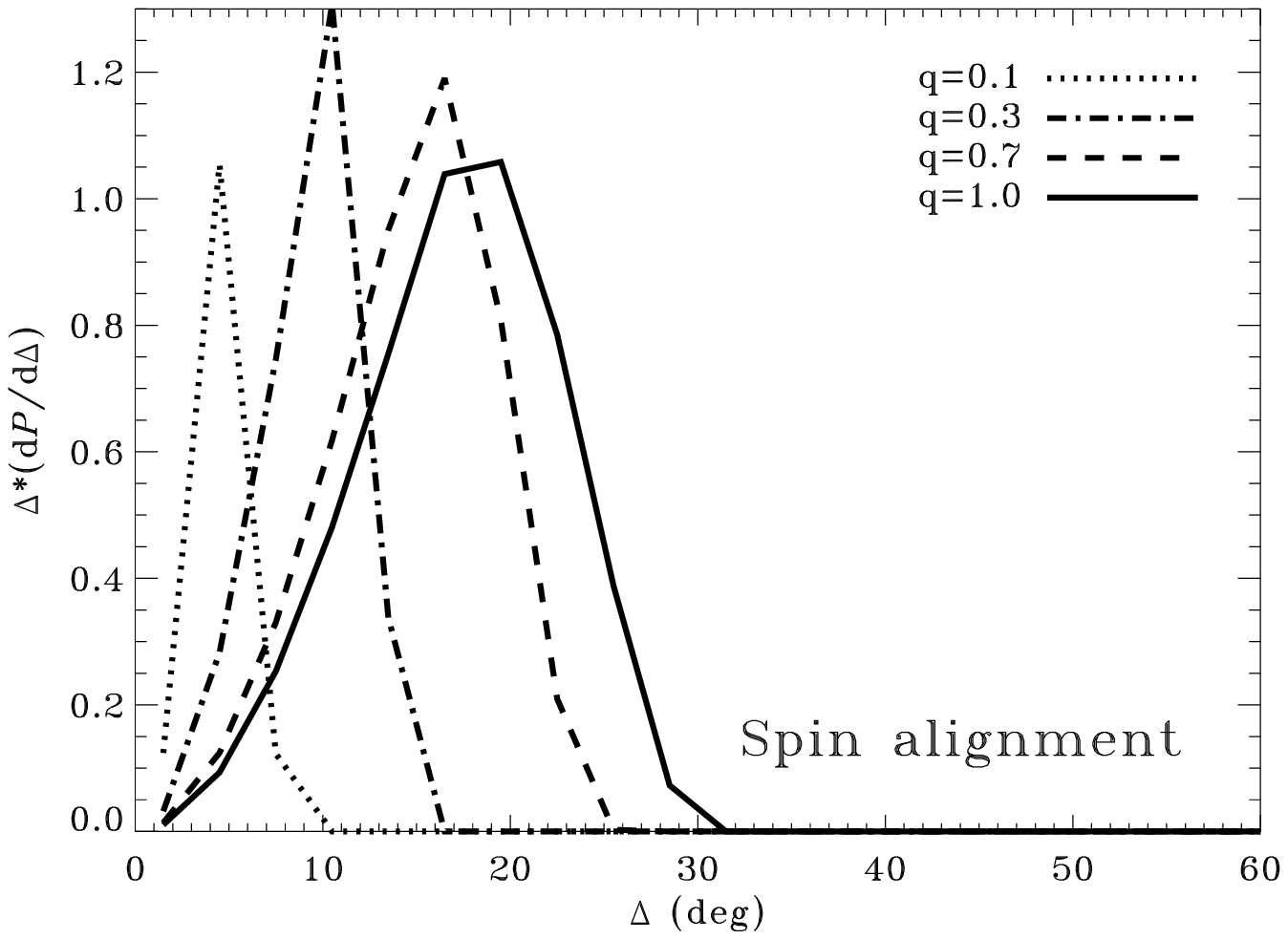}
\caption{Distribution functions of intrinsic spin flip angle $\Delta$
  for different SMBHB mass ratios, assuming random initial spin orientation 
  (left panel) or initial spin alignment within $30^{\circ}$ 
  relative to the binary orbital angular momentum (right panel). 
  Spin-flip angles in the mergers with larger
  mass ratio generally have broader distributions and the equal-mass
  mergers give the largest possible spin flip angle. When the initial
  spin orientations are random (aligned), typical spin flip angles are
  $\Delta \sim 30^{\circ}$ ($\sim19^{\circ}$) for $q=1$, $\Delta \sim
  25^{\circ}$ ($\sim17^{\circ}$) for $q=0.7$, and $\Delta \sim 
  19^{\circ}$ ($\sim11^{\circ}$) for $q=0.3$. While for minor mergers
  of $q=0.1$, the spin flip angles narrowly distribute with ${\rm
  FWHM(\Delta)} \simeq 7^{\circ}$ ($\simeq3^{\circ}$) and a peak at
  $\Delta = 12^{\circ}$ ($\sim4^{\circ}$).} 
\label{delta}
\end{figure}

Our results suggest that the merger of SMBHB with a given mass
ratio can lead to apparent jet reorientation angle $\delta$ in a
broad range, which is partially due to the projection effect. In
this section, we show the distribution function of the intrinsic
spin flip angle $\Delta$ We calculate $\Delta$ using the
relation $\cos(\Delta) = {\bf a}_1 \cdot {\bf a}_{\rm fin}/|{\bf
  a}_1||{\bf a}_{\rm fin}|$ and give the results for
typical mass ratios $q = 0.3$, $q = 0.7$, and $q=1$ in the left panel
of Figure~\ref{delta}. In the same panel, we also give the distribution of
$\Delta$ in minor merger $q=0.1$, which is also the lower
mass ratio limit to form spin-flip radio sources with
moderate apparent jet reorientation angles.

We can see that in spin-flip radio sources the typical
spin flip angle corresponding to the maximum
probability increases with mass ratio $q$ from $\Delta=12^{\circ}$ for
minor merger $q=0.1$, $\Delta= 19^{\circ}$ for major merger $q=0.3$,
and $\Delta=25^{\circ}$ for major merger $q=0.7$ to
$\Delta=30^{\circ}$
for equal-mass merger $q=1$. Our results also show that the spin
flip angle $\Delta$ has a broad distribution and the full width
at half maximum FWHM($\Delta$) also increases with SMBHB
mass ratio from ${\rm FWHM(\Delta)} \simeq 7^{\circ}$ for minor
merger $q=0.1$ through ${\rm FWHM(\Delta)} \simeq 13^{\circ}$ for
major merger $q=0.3$ to ${\rm FWHM(\Delta)} \simeq 28^{\circ}$ for
equal-mass merger $q=1$. For equal-mass mergers ($q=1$), 
we find $\Delta\la55^\circ$. The trial solutions with $\Delta>55^\circ$ 
are mostly rejected because that significant spin down of the 
post-merger SMBH would occur.

The maximum spin flip angle for minor
merger $q=0.1$ is about $15^{\circ}$, which is consistent with the
results given in Figure~\ref{qdis} in which no spin-flip radio source
with apparent jet reorientation angle $\delta > 40^\circ$ results from
minor merger with $q=0.1$. However in Figure~\ref{qdis}, some spin-flip
radio sources with apparent jet reorientation angle $15^\circ < \delta
< 40^\circ$ are from the minor merger with $q=0.1$, this is because of
the projection amplification. Comparing Figure~\ref{qdis} and
Figure~\ref{delta}, one can see that the projection amplification
becomes more important for larger apparent jet reorientation
angle. For mergers with $q=0.3$, the maximum intrinsic spin flip angle
is $\Delta < 30^\circ$ in Figure~\ref{delta}, but in Figure~\ref{qdis}
a significant fraction of spin-flip radio sources with apparent jet
reorientation angle $40^{\circ} <\delta <65^{\circ}$ are due to these
mergers and even their contributions to the formation of spin-flip
radio sources with $65^{\circ} <\delta <90^{\circ}$ are not
negligible because of the projection amplifications. The amplification
effect becomes dominant for the formation of spin-flip radio sources
with $65^{\circ} <\delta <90^{\circ}$, because Figure~\ref{delta}
showed that even for equal-mass mergers $q=1$ the probability to have
spin flip angle larger than $60^\circ$ is nearly zero.

{ It has been suggested that the spins of SMBHs 
may be partially aligned via interactions with a gas disk or
via GR precession \citep{bog09,dot09,kes10} prior to the 
binary coalescence. To understand the effects of spin alignment
on the distribution of spin-flip angle, we ran test simulations
assuming that the black hole spins are initially partially aligned within $30^{\circ}$
relative to the orbital angular momentum of SMBHB \citep[also assumed by][]{lou10}. The corresponding 
results are shown in the right panel of Figure~\ref{delta}. We did not consider 
further alignment within $10^\circ$, because such extreme alignment hardly ever 
produces an apparent spin-flip angle greater than $15^{\circ}$,
the minimum angle required to identify spin flip in radio sources.
When spin alignment is taken into account, the peaks of the distribution functions move
systematically to lower $\Delta$ and the FWHMs shrink significantly. The maximum intrinsic
spin-flip angle, $\sim30^{\circ}$, is achieved when $q=1$. These results imply that
in the context of spin-flip model, radio sources with spin-flip angles much greater
than $\sim30^{\circ}$ are strong evidences against spin alignment during SMBHB coalescence.}

\begin{figure}
\plotone{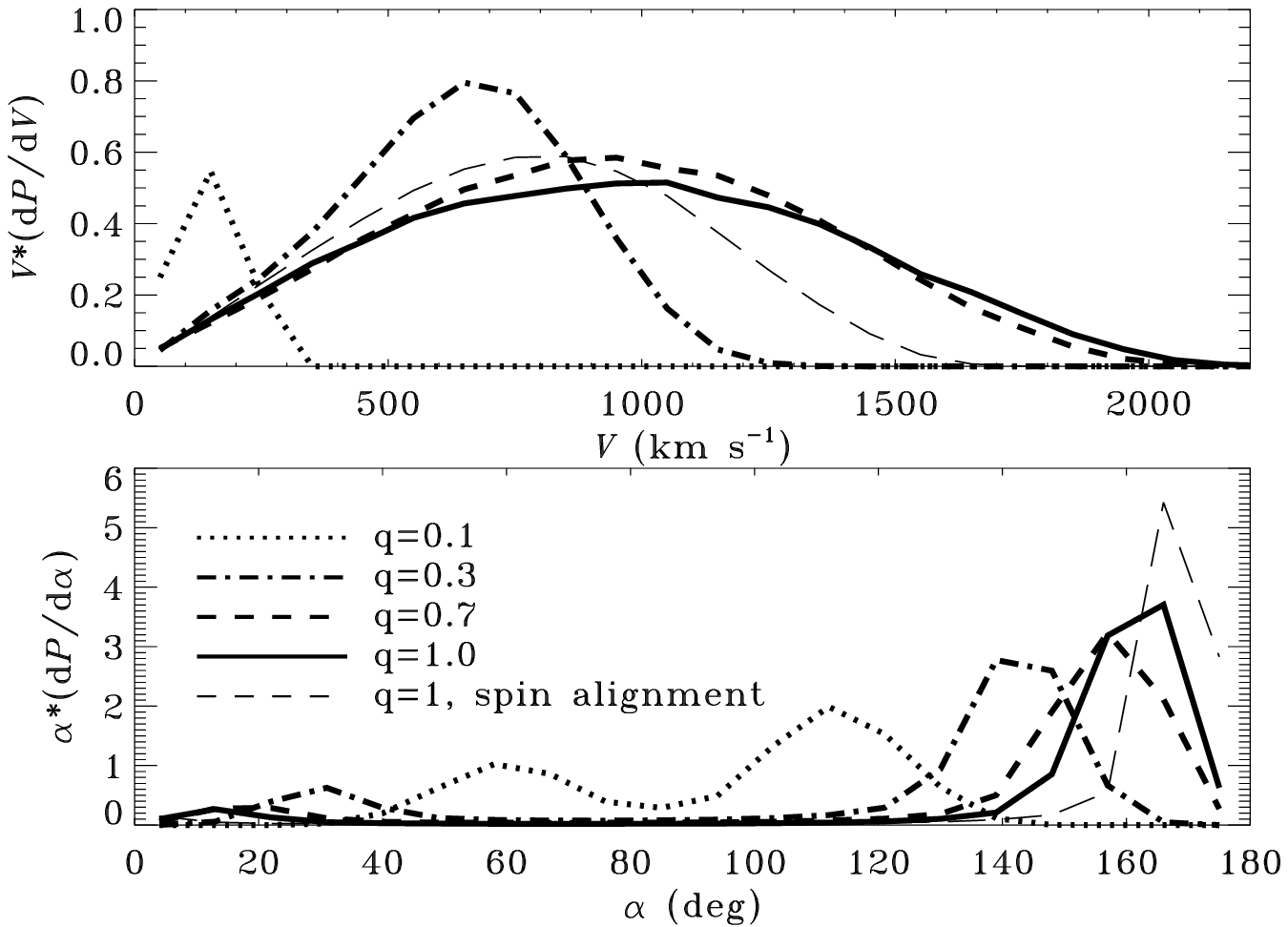}
\caption{({\it Upper}) Distribution functions of recoiling
  velocity $V$ for different black hole mass ratios. For major
  mergers with $q\ga 0.7$, the distributions of recoiling velocity are
  almost identical, ranging from 0 to about $2100\, {\rm km~s}^{-1}$ with
  most possible recoiling velocity $V\sim 900\, {\rm
  km~s}^{-1}$. For smaller mass ratios, $V$ distributes in narrower
  ranges and the fraction with large recoiling
  velocity decreases rapidly. Most mergers have recoiling velocity
  $V\sim 600\, {\rm km~s}^{-1}$ for q=0.3. for minor mergers $q
  \sim 0.1$, recoiling velocity is narrowly peaked at $V\sim 150\,
  {\rm km~s}^{-1}$.
  ({\it Lower}) Distribution functions of the direction angle
  $\alpha$ of recoiling velocity relative to the spin axis of
  post-merger black hole for different mass ratios $q$. For major
  mergers of $q\ga 0.3$, recoiling SMBHs move preferentially along
  the spin axis of post-merger SMBHs within an angle of
  $40^{\circ}$. For major mergers $q\ga 0.7$, most recoiling SMBHs
  moves in a direction of $\alpha \ga 165^{\circ}$. 
  The thin dashed lines in both panels show the results when 
  partial spin alignment
  is considered.}
\label{vdis}
\end{figure}

While the spin axis of jet-ejecting SMBHs reorient during the
merging of SMBHBs, the post-merger SMBHs get kicked due to
the anisotropic radiation of gravitational wave. In the upper panel of
Figure~\ref{vdis}, we give the distribution of the recoiling velocity $V$
in spin-flip radio sources for black holes with random 
initial spin orientation and
typical mass ratios $q=0.3$, 0.7, and 1,
which are, respectively, the typical mass ratios in spin-flip radio
sources with apparent jet reorientation angles $15^\circ < \delta <
40^\circ$, $40^\circ < \delta <65^\circ$, and $65^\circ < \delta <
90^\circ$. Our results show that mergers of SMBHBs with larger mass
ratios could lead to larger recoiling velocities with broader
distributions. For major mergers of $q=0.3$, the recoiling velocities show
a distribution ranging from nearly $0\, {\rm km~s^{-1}}$ to larger
than $1200\, {\rm km~s^{-1}}$ with a most possible velocity  $600\,
{\rm km s^{-1}}$. For major mergers of q=0.7 or q=1, the recoiling velocities $V$
show broader distributions ranging from nearly 0 to larger than
$2100\, {\rm km~s^{-1}}$. The most probable kick velocity is about
$900\, {\rm km~s^{-1}}$, while the typical probability for a
post-merger SMBH to have recoiling velocity larger than $1000\, {\rm
  km~s^{-1}}$ is nearly 50 percent. In Figure~\ref{vdis}, we also
give the results for minor mergers with $q=0.1$, which will lead to
small spin flip of $\Delta \leq 16^\circ$. The kicked SMBHs for
$q=0.1$ have with a very narrow velocity range with ${\rm
FWHM(V)} \simeq 210 \, {\rm km~s^{-1}}$ and the most probable kick
velocity is $150\, {\rm km~s^{-1}}$. Therefore, our results suggest
that the post-merger SMBHs in spin-flip radio sources with larger
apparent jet reorientation angle should have larger recoiling
velocities, in particular for spin-flip radio sources with $15^\circ <
\delta < 40^\circ$ the typical recoiling velocity is about $600\, {\rm
  km~s^{-1}}$, while for spin-flip radio sources with $65^\circ <
\delta < 90^\circ$  the recoiling velocity is typically about $900\,
{\rm km~s^{-1}}$ or larger. If pre-merger spin alignment is
considered, the typical spin-flip angle would be smaller than 
$30^{\circ}$ according to the right panel of Figure~\ref{delta}. 
The typical recoiling velocity would be significantly smaller, as the thin
dashed line shows in the upper panel of Figure~\ref{vdis}.

In addition to the magnitude, it is also important to know the
direction of the recoiling velocity of a kicked SMBH. In the lower
panel of Figure~\ref{vdis}, we give the distribution of the angle
$\alpha$ 
of recoiling velocity relative to the spin axis of post-merger SMBH,
or the orientation of the active jets. Our results suggest that the
kicked SMBHs are inclined to move along the active jets and likely
to avoid moving in the SMBHB orbital plane. Earlier statistical studies 
by \citet{lou10b} on the spin properties of merger remnants showed the similar
trend and attributed the alignment to that both final spin and recoiling
velocity are biased toward (counter-)aligning with the orbital angular momentum
of the pre-merger SMBHB. This explanation implies that if the spin axes
of SMBHs could align with the orientation of orbital angular momentum
prior to coalescence, the bias would be even stronger and the correlation 
between the orientations of final spin and recoiling velocity would be
tighter. The result from our test simulations with partial spin
alignment, which is shown in the lower panel of Figure~\ref{vdis} as
the thin dashed line, supports this implication. For major mergers
with $q> 
0.3$, most kicked SMBHs move in a direction within $40^{\circ}$ with
respect to the final spin axes or the active jets, which is about or
less than the opening angle of the dust torus in AGN unification
model. Therefore, recoiling SMBHs in most spin-flip radio sources do
not move across the dust torus and make direct impact on it, if the
active jets are vertical to the plane of the dust torus. For
equal-mass mergers with $q=1$, nearly all kicked SMBHs move in the
direction $\alpha \sim 15^\circ$ anti-parallel to the final spin
angular momenta. These results imply that if SMBHBs merge in
blazars, the broad emission lines are nearly maximumly Doppler-shifted
for a given recoiling velocity. For minor merger with $q=0.1$, the kicked SMBHs move in a broad range of directions and most of them move in the direction of
about $50^\circ$ relative to the active jets.

%%%%%%%%%%%%%%%%%%%%%%%%

\subsection{Line-of-sight velocities and apparent
  off-center displacements of recoiling AGNs}\label{sec:vl}

The projections of spin flip angles vertical to LOS is the apparent jet
reorientation angles $\delta$ of spin-flip radio sources and can
be given by observations.  The projected components
$V_{\parallel}$ of recoiling velocities along LOS can produce
Doppler-shifted kinematic offsets of broad emission lines with
respect to the narrow emission lines of AGNs or to the spectral lines of
host galaxies, while the other projected component $V_{\bot}$
perpendicular to LOS leads to apparent spatial offsets of AGNs
with respect to the center of the host galaxies. Therefore, in
this section we investigate the Doppler-shifting recoiling
velocity $V_{\parallel}$ and the apparent spatial off-center
displacement $L_{\bot}$ of recoiling SMBHs as a
function of the apparent jet reorientation angle in spin-flip radio
sources. To take into account the deceleration of a recoiling 
SMBH due to dynamical friction against stars in the host galaxy and
dark matter in the halo, we adopted an 1-D analytic model (see
Appendix~\ref{decel}) to calculate $V_{\parallel}$ and $L_{\bot}$ as a
function of $t_e$, the elapsed time since a post-merger SMBH gets kicked.

\citet{arm02} showed that the merger of SMBHB in an AGN would truncate
the inner accretion disk, and \citet{liu03} suggested that the 
jet formation will be interrupted accordingly and
later restart, when the inner disk is refilled after an
interruption timescale $t_{\rm tr}$. In this scenario, the total
elapsed time after the SMBHB coalesces is the sum of
$t_{\rm tr}$ and the life time of the active jets $t_{\rm a}$, i.e.
$t_{\rm e} = t_{\rm a} + t_{\rm tr}$. In X-shaped radio sources, one
can estimate the radiation ages $t_{\rm ra}$ of active radio lobes
using radiation models for relativistic electron plasma in the radio
lobes, or calculate the dynamic lifetime $t_{\rm dy}$ of jets
using the typical length of jets $L_{\rm a}$ and the typical advancing
velocity $V_{\rm hs}$ of hot-spots in the active radio lobes
\citep[e.g.][]{AL87}. We can use either of them as the typical
lifetime $t_{\rm a}$ of active jets. To calculate $t_{\rm tr}$,
we adapted the equation~(10) in \citet{liu03}, $t_{\rm tr} \simeq
4\times10^3~q^{3/5}(M/10^8~\msun)\, {\rm yr}$,  for SMBHBs with 
circular orbit\footnote{ In equation~(10) of \citet{liu03} 
    the factor $f=21.8$ and the resonance number $n = 5$ are adopted 
    for binary orbit with eccentricity $e=0.68$. While, for a
    circular binary orbit which we assumped here, $f=1$ and $n=2$.}
and assumed a viscous parameter of $0.1$ 
and  an opening angle of $0.01$ for a standard thin accretion 
disk. The resulting interruption timescale of jet formation for $M
\la 10^9 \msun$ is and much shorter than the typical lifetime of the
active jets in a giant radio source (GRS, typical jet size
$\sim100$ kpc and age $\sim 3\times 10^6 {\rm yr}$,
e.g. \S~\ref{sec:sample}). However, in a compact steep-spectrum (CSS,
typical jet size $\sim10$ kpc and age $\sim 10^5 {\rm yr}$) or a 
gigahertz peaked-spectrum (GPS, typical jet size $\la 1$ kpc and
age $\la 10^4 {\rm yr}$) radio source \citep{ode98}, the interruption
time $t_{\rm tr}$ become comparable or greater than the lifetime of
the active jets \citep[assuming $V_{\rm hs}\simeq 0.3c$,][]{ows99},
thus not negligible in the calculation of $t_e$.

Radio sources are classified into radio galaxies and
quasars. In the unification model of AGNs \citep{urr95}, radio quasars
are those radio galaxies that the angles between the relativistic jets
and LOS are small so that the accretion disks and the outer dust 
tori are nearly face on. Therefore, in quasars the broad emission line
regions can be observed directly. However, in radio galaxies, the
accretion disks and dust tori are nearly 
edge-on and the broad emission line regions are obscured by dust
tori. The division viewing angle between radio quasars and galaxies is
about $44.4^{\circ}$. In the following, we calculate the
Doppler-shifting recoiling velocities and apparent spatial offsets
for spin-flip radio galaxies with $44.4^{\circ}<w_2<90^{\circ}$ and
for spin-flip quasars with $10^{\circ} <w_2 <44.4^{\circ}$,
respectively. We do not consider the cases with $w_2 < 10^{\circ}$
because such sources will be identified as blazars due to the close
alignment of jets and LOS.

\begin{figure*}
\plottwo{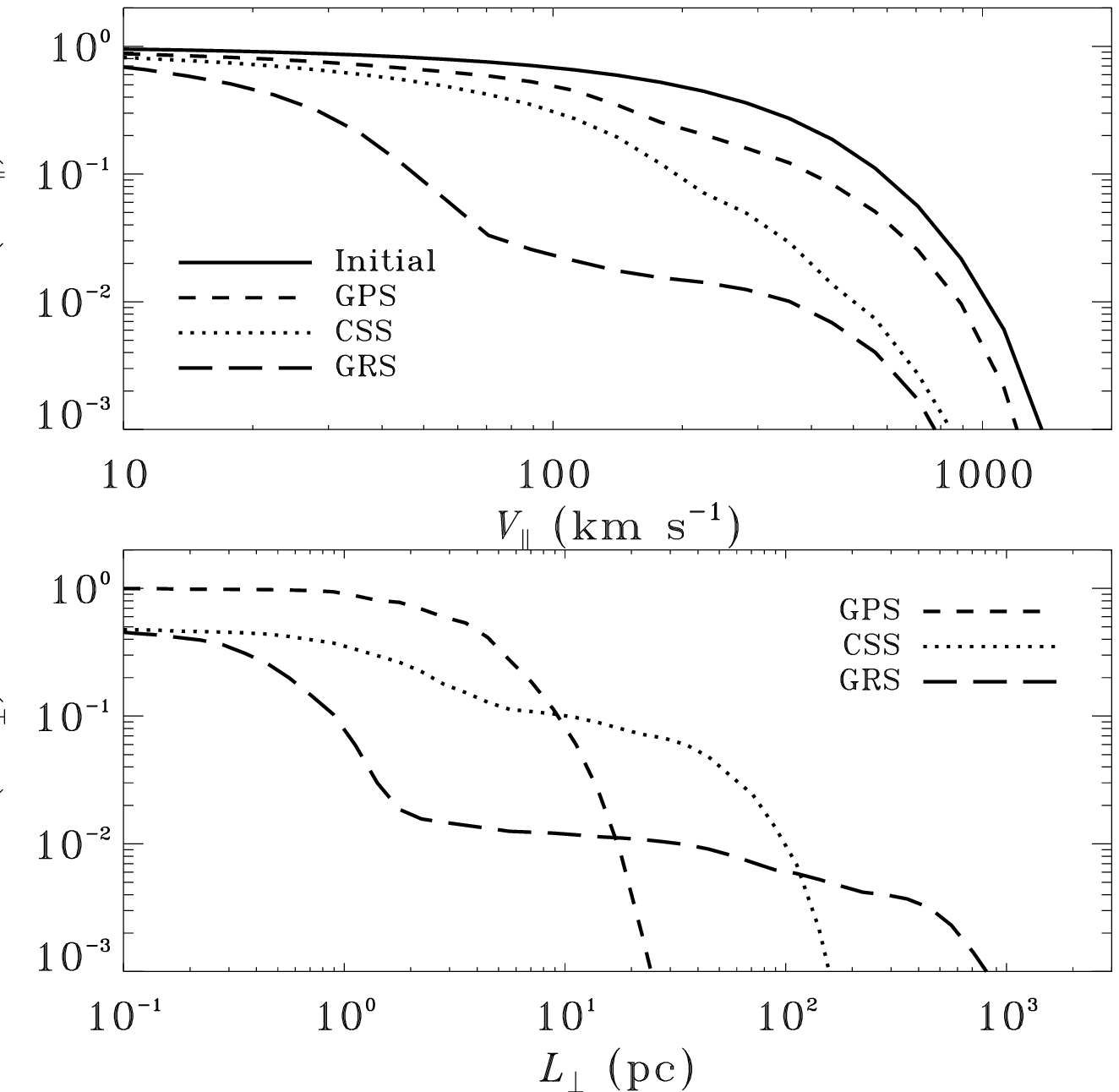}{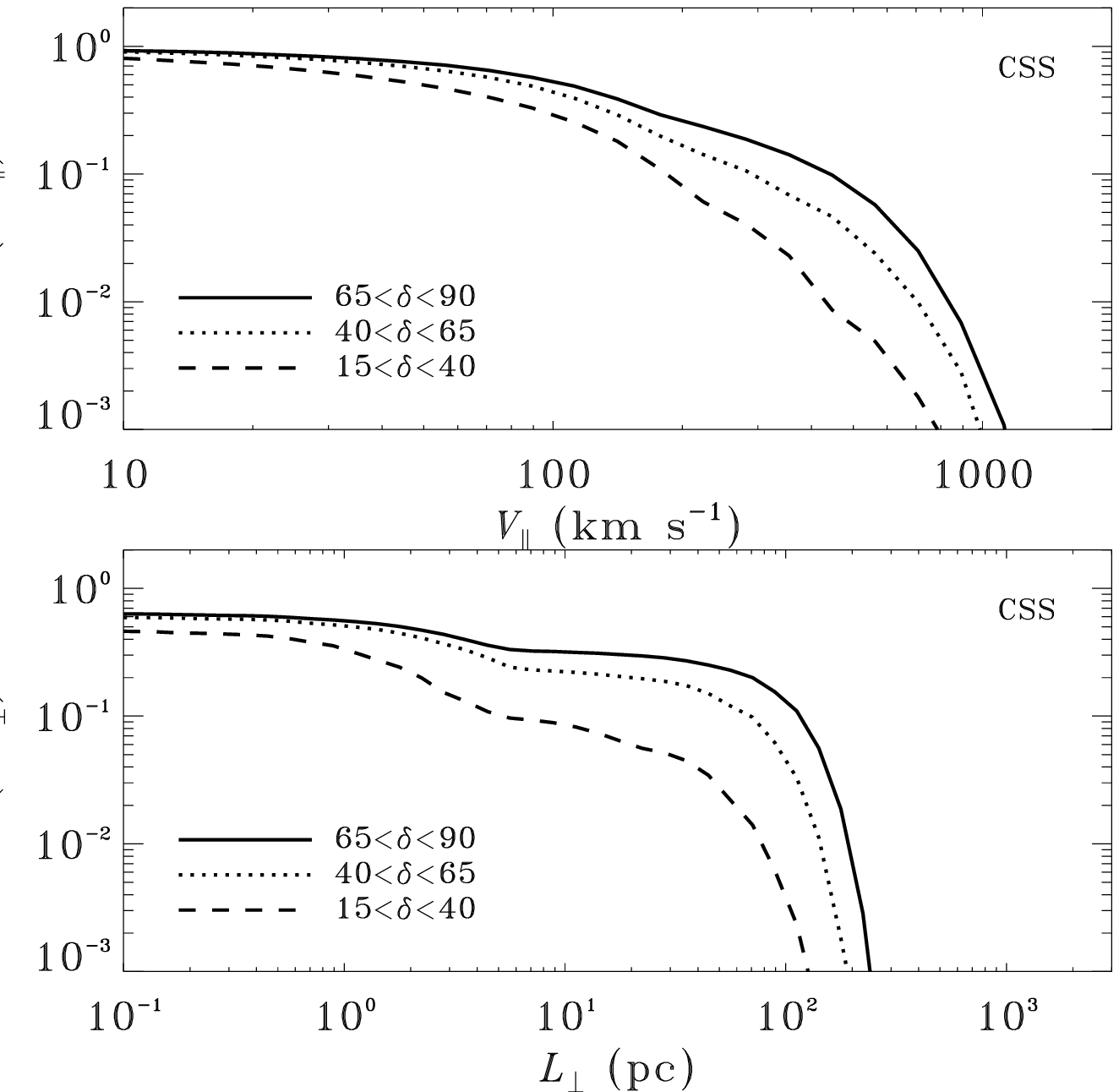}
\caption{{ Cumulative distribution functions of LOS 
    velocity ({\it Upper}) and apparent off-center
    displacement ({\it Lower}), for spin-flip quasars at initial
    kick (with quenched jet formation) and with increasing active jet
    sizes and ages from GPSs to CSSs and finally GRSs ({\it left})
    and for CSS quasars with different apparent spin-flip angles ({\it
    right}). The fiducial black hole mass is $10^8~\msun$ and the
    stellar velocity dispersion of host galaxy} is $200~\kms$.
\label{fig4}}
\end{figure*}

The upper panels of Figure~\ref{fig4} show the cumulative distributions of
the LOS velocity $V_{\parallel}$ in spin-flip quasars. In the
calculation, we assumed a typical black hole mass of $10^8~\msun$ 
and stellar velocity dispersion of host galaxy $200~\kms$. In the left
panel, the solid curve refers to the initial kick velocity immediately after 
SMBHB coalescence, and the other curves refer to quasars with
increasing jet sizes and ages, {namely, GPSs, CSSs, and GRSs. 
With the increase of time since kick and the transition of quasar
types from GPSs to CSSs and finally GRSs, the  
deceleration of recoiling SMBH due to the dynamical 
friction against stars in the host galaxy and dark matter in the halo
becomes more significant, so both the maximum $V_{\parallel}$ and the
probability for $V_{\parallel}>10~\kms$ decreases. For GRSs, the bump
of the cumulative distribution function at $V_\parallel<70~\kms$ is
contributed mainly by the SMBHs in the long-term oscillatory phase II
\citep{gua08}. In the right panel, we show distribution functions of
$V_\parallel$ for CSS quasars with three different apparent spin-flip
angles, i.e., $15^{\circ} < \delta<40^{\circ}$, $40^{\circ} < \delta <
65^{\circ}$, and $65^{\circ} <\delta <90^{\circ}$. The first
distribution function for $15^{\circ} < \delta<40^{\circ}$ resembles
the distribution function for all CSSs shown in the left panel,
indicating that the majority of the spin-flip CSSs have small
$\delta$. While the latter two distribution functions are
significantly higher than the first one, implying that $V_{\parallel}$
is generally larger for a spin-flip quasar with larger apparent jet
reorientation angle. Current spectroscopy with medium
resolution $R\simeq1800$, such as that from Sloan Digital Sky
Survey, can resolve a Doppler-shift velocity of $\Delta
v\simeq167~\kms$ \citep{era11}. A spectral resolution as high as
$46,000$, such as HIRES/Keck and UVES/VLT \citep{ara98,bal11}, can
even resolve $\Delta v\simeq6.5~\kms$. Therefore, with a
conservative assumption of a spectral resolution of $200~\kms$, the
implications from the upper 
panels of Figure~\ref{fig4} are as follows. Among all quasars showing
spin-flip signatures, the probability of detecting Doppler-shifted
broad emission lines drops from as high as $50\%$ for quasars
with quenched jet formation and activity, to $\la 20\%$ in GPS
and CSS quasars, and to $\la2\%$ in 
GRSs. However, if only quasars with $\delta>40^{\circ}$ are
considered, the detection probability would rise by several times up
to an order of magnitude.  The above results also suggest that
spin-flip quasars hardly ever have LOS velocities greater than
$2000\,{\rm km~ s^{-1}}$. 

In the lower panels of Figure~\ref{fig4}, we give the cumulative
distribution functions of $L_{\bot}$, the apparent off-center
displacement, for quasars with different typical jet sizes ({\it
left}) and with different apparent spin-flip angles ({\it right}). 
As the jet size grows and deceleration proceeds, the maximum
$L_{\bot}$ increase while the probability for $L_{\bot}>1$ pc
decreases. The probability for a spin-flip quasar to have $L_{\bot}>2$
kpc is negligible, therefore the recoiling quasar is confined in its
host galaxy. We can also see that for quasars with larger $\delta$,
the probability of $L_{\bot}>1$ pc is generally greater. 
To detect an off-center quasar, one needs to observe the angular 
offset between the recoiling SMBH and the center of its host galaxy.
Using the images obtained with {\it Hubble Space Telescope (HST)} of
spatial resolution about 100 milli arcsec (mas) and a relative
astrometric accuracy $\sim 12$ mas, \citet{bat10} recently detected an
angular displacement of $\sim 100$ mas between the SMBH and the
photo-center center of host galaxy in M87. The detection implies that
the relative angular astrometric accuray of a telescope may be more
important in observing off-center displacements of recoiling SMBHs. 
The current large telescopes, e.g. the Keck II telescope and {\it Very 
Large Telescope (VLT)} equipped with adaptive optics (AO) systems, can
achive the highest angular resolution $\sim 20 \, {\rm mas}$ and
relative astrometric accuracies $\sim 0.2 \, {\rm mas}$ at
infrared wavelength \citep{wiz00,lu10}, while future astrometric
interferometer 
missions, such as Gaia, are designed to achieve sub-mas resolution
\citep[e.g.][]{pop11}. At a distance of $100$ Mpc (1 Gpc), an angular
displacement of $0.1$ mas corresponds to a physical scale of 0.05
  (0.5) pc. The lower panels of Figure~\ref{fig4} indicates that the
probability with large telescopes with AO systems or Gaia
detecting off-centered quasars among the spin-flip quasar sample is
$\ga 50\%$ ($\ga 30~\%$) for 0.05 (0.5) pc. In the GPS
subsample, especially, the detection probability is
$95\%$ both for 0.05 pc and 0.5 pc.

\begin{figure}
\plottwo{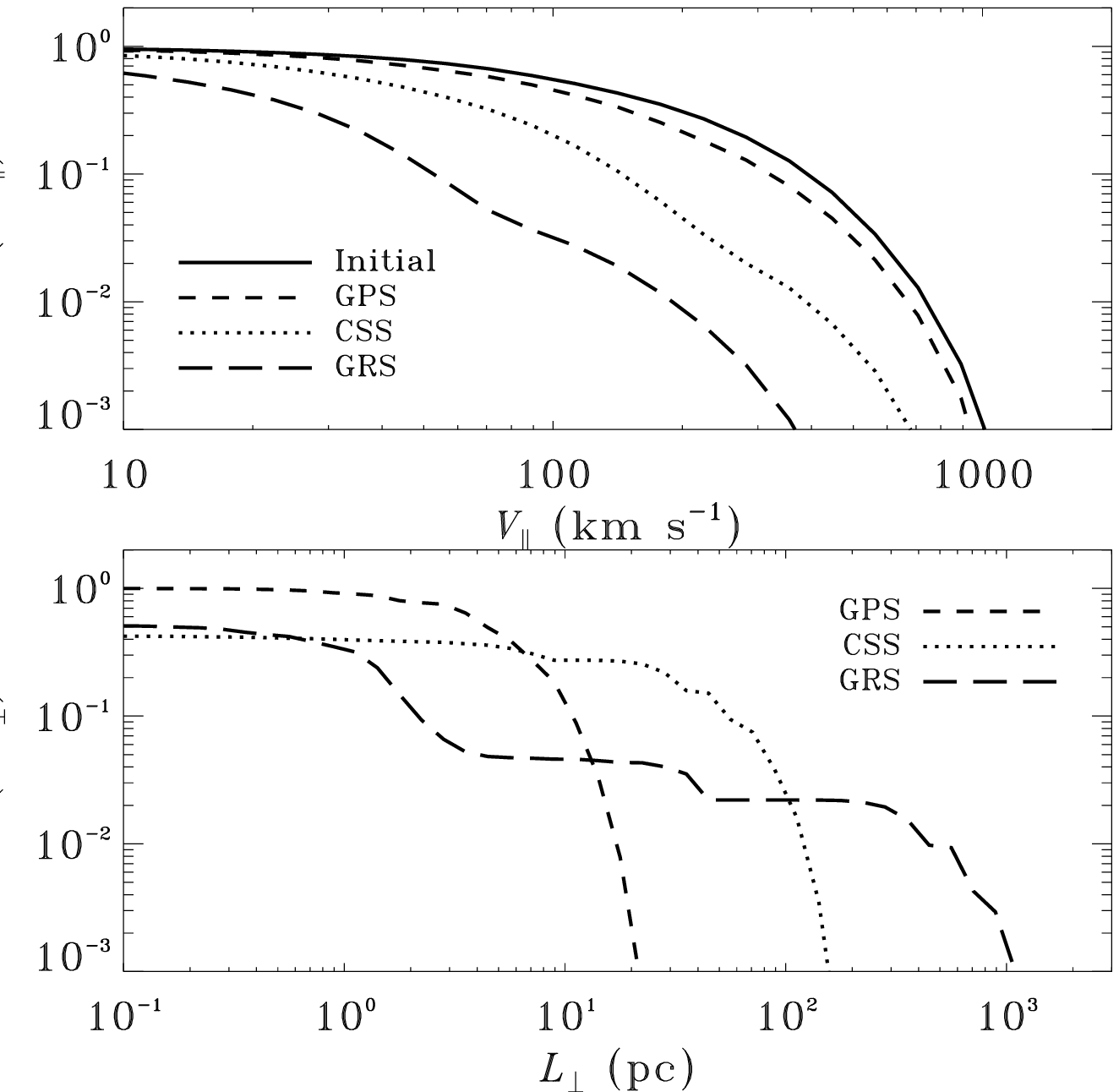}{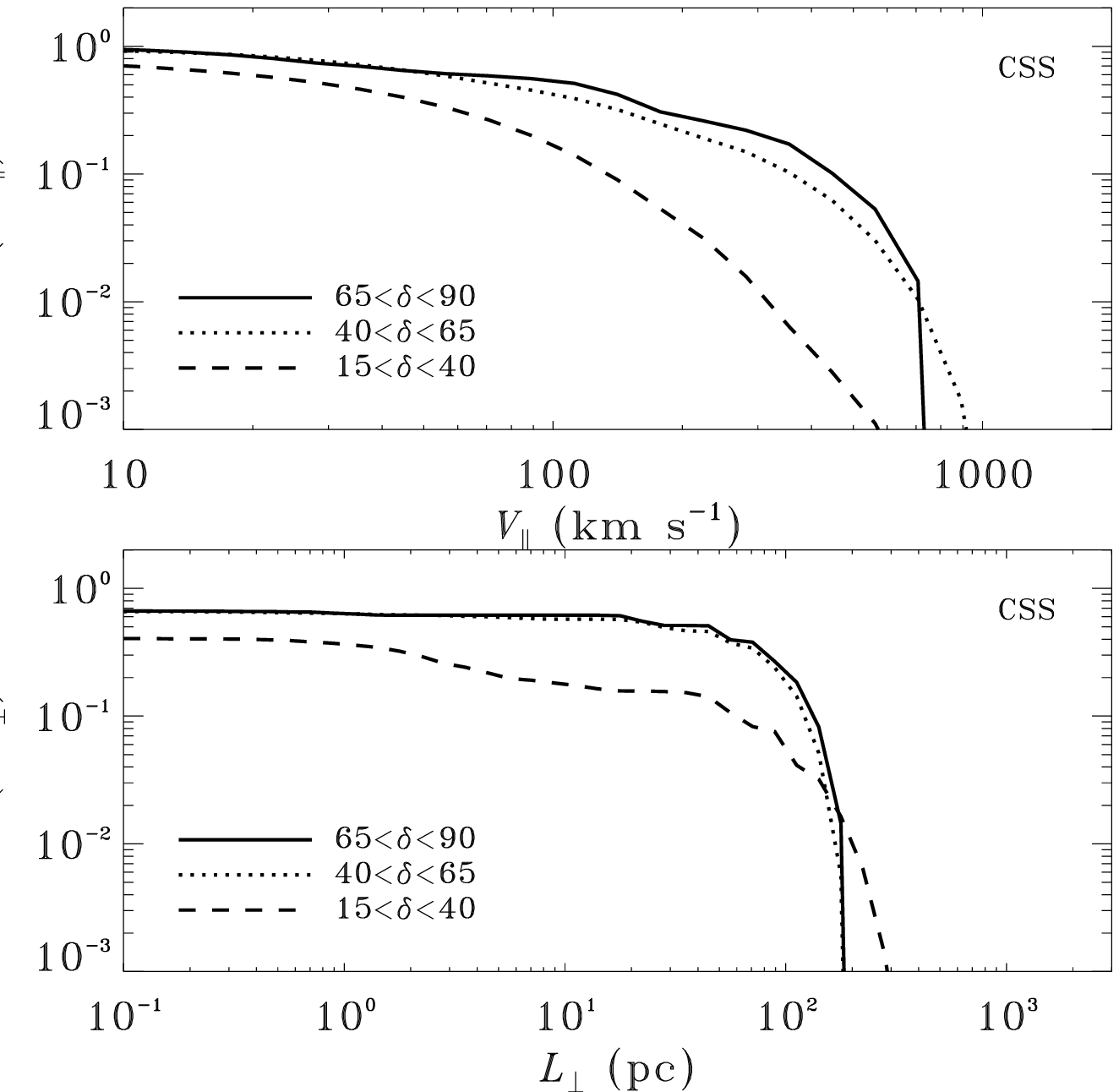}
\caption{{ The same as Figure~\ref{fig4} but for spin-flip radio galaxies,
which have $44.4^{\circ}<w_2<90^{\circ}$.}}
\label{fig5}
\end{figure}

Figure~\ref{fig5} shows the same distribution functions as in
Figure~\ref{fig4}, but for spin-flip galaxies. The distribution
functions show similar dependence on typical jet size and apparent
spin-flip angle as that in spin-flip quasars. However, due to the
projection effect, the maximum and mean values of $V_\parallel$ 
for spin-flip galaxies are smaller than those for spin-flip quasars,
and the maximum and mean values of $L_{\bot}$ are slightly greater.

The above results are derived assuming a black hole mass of $10^8~\msun$. The
corresponding kick velocity for a recoiling SMBH to escape to the effective radius of the
host galaxy  
is about $1600~\kms$. For systems with larger (smaller) black holes, the escape 
velocity becomes higher (lower). As a result, given $t_e$, the probability of detecting either Doppler-shifted broad emission lines or off-center AGN displacement would decrease (rise). We did not consider spin alignment in this subsection, which (if taken into 
account) would reduce $v_k$ and the possibilities of detecting appreciable 
$V_\parallel$ or $L_\bot$. Since SMBHB coalescence with spin alignment
hardly ever produces any radio source with $\Delta>40^\circ$, radio sources showing 
large apparent spin-flip angles $\delta>40^\circ$ are consistent with random spin
orientation, and for these sources our conclusions are unchanged.
%---
\subsection{Truncation of dust torus and dust poor AGNs}

The dust torus around central SMBH is an essential component in the
unification model of AGNs. The inner edge of dust torus, $r_{\rm
  tor,in}$, is most probably determined by the dust sublimation radius
\begin{equation}
  r_{\rm tor, in} \ga 0.4 \left({L_{\rm bol} \over {10^{45} {\rm erg\
	s}^{-1}}}\right)^{1/2} \left({1500 {\rm K} \over {T_{\rm
	sub}}}\right)^{2.6}~ {\rm pc}
\label{eq:rtorus}
\end{equation}
\citep[e.g.][]{nen08}, where $L_{\rm bol}$ is the AGN bolometric
luminosity, and $T_{\rm sub}$ is the dust sublimation
temperature. The outer radius of dust torus, $r_{\rm tor, out}$, is
poorly determined but likely not much larger than $r_{\rm tor, out}
\sim 20 r_{\rm tor, in}$ \citep[e.g.][]{eli07}.
Inside the dust torus is the broad emission line region, the size of
which can be given with the empirical scaling relation
\begin{equation}
r_{\rm BLR} = 1.9\times 10^{-2} \left[{\lambda
    L_{\lambda}{\rm(5100\AA)} \over {10^{44} {\rm erg\
	s}^{-1}}}\right]^{0.69} \, {\rm pc} ,
\label{eq:blr}
\end{equation}
where $L_{\lambda}{\rm(5100\AA)}$ is the AGN luminosity at $\lambda =
5100{\rm\AA}$ \citep{kas05}. The dichotomy of Type II and Type I quasars is
then caused by whether or not LOS toward the broad line regions of AGNs is blocked
by the dust tori.

\citet{KM08a} showed that in a large fraction of recoiling quasars the
matter in dust tori would become unbound to the recoiling SMBHs and
the Type II quasars obscured by the dust tori would transform to
unobscured Type Is. When a SMBH gets a kick velocity $V$, the matter
orbiting the SMBH at radius $r \ga r_{\rm k}$ becomes unbound and only
those with $r\la r_{\rm K}$ remains bound to the recoiling SMBHs.
The critical radius is
\begin{equation}
  r_{\rm k} \simeq {GM \over V^2} \approx 0.43
  \left({M \over{10^8 M_\odot}}\right)
  \left({V \over {10^3 {\rm km\ s}^{-1}}}\right)^{-2} ~{\rm pc} ,
\label{eq:rkick}
\end{equation}
where $M_\odot$ is the solar mass and $M$ is the mass of kicked SMBH
after merger. If $V$ is larger than a critical value $V_{\rm cr}$
which satisfies $r_{\rm k} = r_{\rm tor, in}$, all the matter in dust
torus becomes unbound, therefore the recoiling quasar must be ``dust
free'', and the initial radio galaxy will become an off-center
quasar. If the recoiling velocity $V$ is not only larger than $V_{\rm
  cr}$ but also
\begin{equation}
V_{\rm wl} \simeq 4.8\times 10^3 \left({M \over{10^8
    M_\odot}}\right)^{1/2} \left[{\lambda
    L_{\lambda}{\rm(5100\AA)} \over {10^{44} {\rm erg\
	s}^{-1}}}\right]^{-0.35} \, {\rm km\ s}^{-1} ,
\label{eq:wl}
\end{equation}
a part of the matter in broad line region may also become unbound and
the recoiling quasar will become a weak emission line object.
However, if the recoiling
velocity is moderate and $V \la V_{\rm dp} \sim V_{\rm cr} (r_{\rm
tor,in}/r_{\rm tor,out})^{1/2}$, the critical radius
$r_{\rm k}$ is larger than $r_{\rm tor, in}$ but smaller than
$r_{\rm tor, out}$. Then part of the dust torus of the recoiling
AGN will become unbound, and the object will become ``dust poor''.

For a typical radio source of $M = 2\times 10^8M_\odot$,
$T_{\rm sub}=1500 K$, $f_{\rm tor}=r_{\rm tor,out}/r_{\rm tor,in}\sim20$, 
and $L_{\rm bol}=10^{45}\, {\rm erg~s^{-1}}$ or
$\lambda L_{\lambda}{\rm(5100\AA)} \simeq {10^{44} {\rm erg~s}^{-1}}$,
the critical recoiling velocity for the dust torus to be completely
truncated and the object to become ``dust free'' is $V_{\rm cr} =
1466\, {\rm km~s^{-1}}$. Meanwhile, the critical velocity for the dust
torus to be truncated partially and the object to become ``dust poor''
is $V_{\rm dp} \sim  300\, {\rm km~s^{-1}}$. Our calculations
  show that the probabilities for such a typical
 radio source to have velocity larger than the critical
velocity $V_{\rm cr}$ are approximately $0.9\% $ for $15^{\circ}
  <\delta <40^{\circ}$, $11\% $ for $40^{\circ} <\delta
  <65^{\circ}$, and $10\% $ for $65^{\circ} <\delta
  <90^{\circ}$. Therefore, only
about $10\%$ of spin-flip radio sources with $\delta>40^{\circ}$ will become ``dust free'' quasars or broad
emission line radio galaxies if broad line regions are present.
However, the critical velocity $V_{\rm dp}$ is at least two times 
smaller than the most-probable velocity in spin-flip radio 
sources (c.f. Figures~\ref{qdis}
and \ref{vdis}). Our calculations indicate that the
probabilities for a typical spin-flip radio source to become ``dust
poor'' are approximately $72\% $ for $15^{\circ} <\delta< 40^{\circ}$,
$85\% $ for $40^{\circ} <\delta <65^{\circ}$, and $85\% $ for
$65^{\circ} <\delta <90^{\circ}$. Therefore, most spin-flip radio
sources should be ``dust poor''. To become a weak emission line
quasar, the recoiling velocity in a typical spin-flip radio source
should be $V \ga V_{\rm wl} \simeq 6.7\times 10^3 \, {\rm km~s^{-1}}$,
which is larger than the maximum 
possible recoiling velocity produced during the merger of SMBHB with
circular orbit. Since the bolometric luminosity $L_{\rm bol}$ should not
greatly exceed the Eddington limit $\sim10^{46}(M/10^8~\msun)~{\rm
  erg~s^{-1}}$, equation~(\ref{eq:wl}) imposes a lower limit on
$V_{\rm wl}$ for a given black hole mass, which is 
$\sim9.5\times10^2~f_{\rm bol}^{0.35}(M/10^8~\msun)^{0.15}~\kms$,
where $f_{\rm bol}=L_{\rm bol}/[\lambda L_{\lambda}{\rm(5100\AA)}]$
is the bolometric correction factor typically in the range $5<f_{\rm bol}<20$
for AGNs. For less massive recoiling black holes, this lower limit
would be smaller 
and more easily achievable. Therefore, weak-line AGNs, if exist, 
preferentially reside in AGNs with small black holes 
($M\sim10^6~\msun$) and high luminosity.

%%%%%%%
\section{Recoiling SMBHs in a sample of X-shaped radio objects}
\label{sec:sample}

In section~\ref{sec:results}, in order to investigate the
distributions of $V_{\parallel}$ and $L_{\bot}$ for spin-flip radio
sources, we have to specify the value or make assumptions for the
distributions of the unknowns, such as $q$, ${\bf
a}_1$, ${\bf a}_2$, ${\bf a}_{\rm fin}$, $t_{\rm a}$, $\delta$, $w_2$,
etc. However, for a real X-shaped radio object, constraints on
the distributions of several unknowns can be made by
observations. For example, according to section~\ref{sec:mratio}, the
formation of X-shaped radio feature requires mass ratio $q\ga 0.1$ and most
probably $q\ga 0.2$. The fact that X-shaped sources are radio loud
AGNs implies $a_1 > 0.9$ and $a_{\rm fin} > 0.9$. In this
section, we study the conditional probability distributions of
$V_{\parallel}$ and $L_{\bot}$ under more observational constraints
for a sample of real X-shaped radio objects.

%------
\subsection{The sample of X-shaped radio objects}

\citet{mez11} studied a sample of 29 X-shaped radio objects,
which is drawn from the list of 100 X-shaped radio source candidates
\citep{che07}. They estimated for their sample objects 
the dynamic ages of the active radio lobes and the central
black hole masses.  The central black hole masses are
computed with the empirical relationships
either between SMBH mass and the stellar velocity dispersion of host galaxy,
or between the width of broad emission line  and optical luminosity
$\lambda L_{5100 {\rm\AA}}$ at $\lambda =5100 {\rm\AA}$. \citet{lan10}
studied the optical spectra of another sample of 
X-shaped radio objects and estimated the viewing angles $w_2$ of
the active jets, based on the relationship between the viewing
  angles and the rest frame equivalent widths of the
narrow emission lines \ion{O}{3}~$\lambda 5007$ and
\ion{O}{2}~$\lambda 3727$. We cross-identified the two samples of
X-shaped radio objects and obtained 23 objects to constitute our
sample of X-shaped radio objects for further study.

We read the viewing angles $w_2$ for the 23 X-shaped radio sources
from the Figure~1 of \citet{lan10} in which they divided the objects
into four groups according to $0^{\circ} < w_2 < 15^{\circ}$,
$15^{\circ} < w_2 < 35^{\circ}$, $35^{\circ} < w_2 < 60^{\circ}$, and
$60^{\circ} < w_2 < 90^{\circ}$. We will follow the division of the
objects because of the large uncertainties in the estimations of the
viewing angles. In \citet{lan10}, a few of our sample objects
were classified into the group of $60^{\circ} < w_2 < 90^{\circ}$, 
{ while upper limits, $w_2 \la70^{\circ}$, were presented in their plot.
Therefore, we assign these objects in a range of 
$45^{\circ} <w_2 <70^{\circ}$ in our calculations.}

We measured the apparent jet reorientation angles $\delta$, the angle
between the orientations of active lobes and wings, of the 23 X-shaped
radio objects from the radio images given by \citet{lal07} for 3C192,
4C+32.25, 4C+48.29, 1059+169, and 3C223.1, by \citet{wan03} for
4C+01.30, and by \citet{che07} for the rest 17 objects. To do the
measurement, we have to define the orientations of the active lobes
and the wings. The orientation of active lobes is defined with the
line passing through the center of the host galaxy and the
hot spots inside the active radio lobes. However, it is very difficult
to define the orientation of wings because they are very
wide, diffusive, and without hot spots inside or central bright plasma
jets along them. To overcome this problem, we define two orientations
of the edges of a wing, both of which starting from the galactic
center but one along the most furthest side (contour) and the other
along the closest side (contour) of the wing with respect to the
active jets. With the two
orientations, we then measured two angles $\delta_1$ and $\delta_2$
between the orientations of active lobes and the two edges of wings,
respectively, and took the mean value of $\delta_1$ and $\delta_2$ as
the value of apparent jet reorientation angle $\delta$. Because of the
limited image quality, the obtained values for $\delta$ have very
large uncertainties. We take $15^{\circ}$ as the typical uncertainty
for the measurement of $\delta$, but for some objects the uncertainties
are much larger. In particular, the uncertainties of $\delta$ for the
objects J0813+4347, J1043+3131, J1111+4050, and J1210+1121
are extremely large, therefore these objects will not be included in
our further calculations. Finally, our sample of X-shaped radio
sources consists of 19 objects which are listed in
Table~\ref{tab:sample}.

In Table~\ref{tab:sample}, Column~1 lists the IAU names of the objects
based on the J2000.0 coordinates and other common catalog names, and
Column~2 gives the spectroscopic identification as well as redshift, G
for radio galaxies and Q for radio quasars. Columns~3, 4, 6, 7, and 8
are, respectively, the masses of central SMBHs, the optical
luminosities of AGNs at $5100 {\rm\AA}$, the viewing angles $w_2$ and
the dynamic ages $t_{\rm a}$ of active lobes, and the references 
{($V_{\rm hs}=0.1c$ for GRSs is assumed in the reference literatures)}. In
Column~5, we give the angles $\delta$ between the active lobes and
wings and the uncertainties measured in this paper. Column~9
gives the critical velocities in unit of $100~\kms$ for AGNs to become
dust-free, calculated according to $r_k=r_{\rm tor,in}$ and assuming
 $f_{\rm bol}=10$. Column~10 gives the critical velocities for AGNs to
become dust-poor, calculated according to $r_k=r_{\rm tor,out}$ and
assuming that $f_{\rm tor}=20$. Column~11 is
the viewing angles $w_1$ of the wings. Because $w_1$ cannot be
constrained observationally and could be any value in the range of
$20^{\circ} \la w_1 \leq 90^{\circ}$, we considered three possible
bins, namely ($20^{\circ}$, $40^{\circ}$), ($40^{\circ}$,
$60^{\circ}$), and ($60^{\circ}$, $90^{\circ}$).

%-------
\subsection{Spectral kinematic offsets and apparent spatial
  displacements}

To compute the projected recoiling velocities along and vertical to LOS using
Equations~(\ref{equ:spin_par})-(\ref{equ:recoil_velp}), we have to
know the mass ratio $q$, spin vectors ${\bf a}_1$ and ${\bf a}_2$, and
the LOS vector $\bf{e_s}$ for each object. From observations
we can estimate only the quantities $w_1$, $w_2$, and $\delta$, and
may assume $a_1 >0.9$ and $a_{\rm fin} >0.9$
as in section~\ref{sec:results}. For the mass ratio $q$, the results
given in section~\ref{sec:mratio} suggest that $q > 0.1$ and most
probably $q > q_{\rm min} \approx 0.2$.  Therefore, we may put a lower
limit for $q$ in our simulations. Our simulations showed that
different lower limits of mass ratio $q$ do not change the conclusions
significantly. We do not have any further knowledge of $q$, ${\bf
  a}_1$, ${\bf a}_2$ and $\bf{e_s}$. Therefore, we assumed
possible distributions of the quantities and used Monte Carlo
simulations to generate their values. For simplicity, we assumed
flat probability distributions of the parameters $q$ with $q_{\rm min}
\leq q \leq 1$, $a_1$ with  {$0.9 \leq  a_1 \leq 0.998$}, and $a_2$
with $0\leq a_2 \leq 0.998$. We also assumed  random orientations of
spin vectors ${\bf 
a}_1$, ${\bf a}_2$, and the LOS vector $\bf{e_s}$. Given ${\bf
a}_1$, ${\bf a}_2$, and $\bf{e_s}$, we calculated
$w_1$, $w_2$, and $\delta$ according to the relevant formula in
section~\ref{sec:method}, and then we used the observed $w_1$, $w_2$,
and $\delta$ given in Table~\ref{tab:sample} to select viable
solutions for (${\bf a}_1$, ${\bf a}_2$, $\bf{e_s}$). Finally,
{ given $q$, $M$, and $t_a$, we derive $t_e$ according to the
scheme explained in \S~\ref{sec:vl}, and then compute
the decelerated LOS velocity $V_\parallel$ and apparent off-center displacement $L_\bot$, using the 1-D analytical model
described in Appendix~\ref{decel}.}

As in section~\ref{sec:results}, Monte Carlo simulations are carried
out for each of the 19 X-shaped radio objects in our sample. 
We derived the probability distributions of Doppler-shifting recoiling
velocity and apparent spatial off-center displacements in the objects,
based on the $10^7$--$10^9$ sets of trial solutions to the
equations and excluding those which are inconsistent with the
observational constraints. The final results are obtained with
about $10^5$-$10^6$ trial solutions. In our simulations for the
objects J1101+1640, J1210-0341, J1327-0203, and J1444+4147, no trial
solution of the equations~(\ref{equ:spin_par})-(\ref{equ:recoil_velp})
is found to fit all the observational conditions. This may imply
that either our assumptions for the physical quantities are
invalid or the spin-flip is not applicable for these
objects.  For the rest 15 objects in our sample,
final solutions to the equations can be obtained 
at least for one of the three bins of viewing angle, namely
$20^{\circ} < w_1 < 40^{\circ}$, $40^{\circ} < w_1< 60^{\circ}$, and
$60^{\circ} < w_1< 90^{\circ}$. The cumulative distributions of the
LOS velocity $V_{\parallel}$ and the apparent
offset-center displacement $L_{\bot}$ for the 15 objects are given in
Figs.~\ref{fig6}-\ref{fig8}. Based on the cumulative distribution functions,
we then derive the probability of displaying Doppler-shifted broad
emission lines, $P_{v,\parallel}$, and off-center spatial
displacements, $P_{\theta,\bot}$. These two probabilities are given,
respectively, in Columns 12 and 13 of Table~\ref{tab:sample}. The two
values in Column 12, from left to right, correspond to
$V_\parallel>500~\kms$ and $V_\parallel>200~\kms$, while the two
values in Column 13 correspond to angular sizes $\theta_\bot>10$ mas
  and $\theta_\bot>0.1$ mas. 

The probability $P_{v,\parallel}$ for our sample of X-shaped radio sources
is typically smaller than $10\%$, except for J0941-0143, 4C+01.30, 
J1348+4411, and J1614+2817. While $P_{v,\parallel}$ for the former
three object are still smaller than $20\%$, the last object,
J1348+4411, shows an appreciable probability ($\simeq40\%$) of having 
$V_\parallel>200~\kms$. In the case of $P_{\theta,\bot}$, for
$\theta_\bot>10$ mas, two objects have these probabilities greater
than $20\%$, i.e., J0941$-$0143 and J1348$+$4411. When
$\theta_\bot>0.1$ mas is considered, all the 15 objects with viable
solutions have $P_{\theta,\bot}>20\%$. In particular, four objects,
4C+32.25, J0941-0143, J1348+4411, and J1614+2817, have
$P_{\theta,\bot}>50\%$ when $\theta_\bot>0.1$ mas. Therefore,
J1348+4411 would be intriguing object for follow-up observations, in
the sense that it has an appreciable probability to show both
Doppler-shifted broads emission lines and off-center spatial displacement.

%---
\subsection{Truncation of dust torus and dust poorness of
  the sample objects}

{ Given $V_{\rm cr}$ and $V_{\rm dp}$ for each object, 
we also calculated the probability $P_{\rm df}$ of an X-shaped radio
object to have {\it initial} recoiling velocity larger than $V_{\rm cr}$
and the probability $P_{\rm dp}$ to have recoiling velocity 
larger than $V_{\rm dp}$. The 
probability $P_{\rm df}$ is also the probability for a sample
object to have type transition from radio galaxies to quasars
and to become ``dust free''. 

We did not present
the probability $P_{\rm df}$ in Table~\ref{tab:sample}, 
because it is less than $20\%$ for most of the objects except
for the quasar 4C+01.30 and radio galaxy J1348+4411. 
The probability $P_{\rm df}$ for 4C+01.30 to 
have recoiling velocity larger than its critical velocity $V_{\rm cr}
\approx 700~(1000)~{\rm km~s^{-1}}$ is about $50\%$ ($30\%$),
{ assuming $f_{\rm bol}=20~(5)$. 
In particular, for $40^\circ<w_1<60^\circ$
(so that torus is ``edge-on'' in the initial configuration) the resulting probability
for this object to be dust-free is about $34\%$ ($60\%$). }
This result implies that the AGN could be obscured before binary black 
hole merger, but the SMBH may have
left behind the entire dust torus due to recoil and 
now transformed into a quasar.
While for J1348+4411, its low critical velocity 
$V_{\rm cr}\approx 560~(800)~{\rm km~s^{-1}}$ suggests that 
$P_{\rm df}\simeq63 \%$ ($45\%$), thus the object may
transform to a quasar or broad emission line radio galaxy.}
However, the probability for the object to remain as a radio galaxy is
{ 37\% (55\%)} and not negligible. Therefore, non-detection of broad
emission lines does not imply that the SMBHB spin-flip model is
completely excluded for this object.

The probability $P_{\rm dp}$ of an X-shaped radio object to have recoiling 
velocity larger than $V_{\rm dp}$ and to become ``dust poor'' is given
in Column (14) of Table~\ref{tab:sample}, assuming that $f_{\rm bol}=10$
and $f_{\rm tor}=20$. The upper and lower indices show the upper and
lower limits to $P_{\rm dp}$ when a range of bolometric correction
factor $5<f_{\rm bol}<20$ and torus size factor $10<f_{\rm tor}<40$
are considered. Among the 15 objects with valid solutions, 12 have
  $P_{\rm dp} > 50\%$, 
and six of them even have $P_{\rm dp} \ga 80\%$. The upper limit of 
$P_{\rm dp}$ is greater than $70\%$ for almost all objects,
and even greater than $95\%$ for 4C+01.30 and J1348+4411.
The poorness of dust in AGN can be observationally tested with the aid
of IR satellites such as {\it Spitzer} \citep{jia10,hao10,hao11}. 
Therefore, the six objects with $P_{\rm dp} \ga 80\%$, namely,
  3C192, 4C+32.25, J0941-0143, 4C+01.30, J1207+3352, and J1348+4411,
  would be interesting targets for IR follow-up observations. We note
  that for 3C192, 4C+32.25, and J1207+3352, the probabilities for the
  off-center displacement $L_\bot$ to be smaller than the typical size
  of dust torus are $>90\%$. The SMBHs in these objects are likely 
settling back in the galaxy nuclei and may become obscured by a
new-forming, next-generation dust torus. 

%---
\subsection{Comments on individual objects}
\label{sec:indiv}

Figures~\ref{fig6}-\ref{fig8} show the cumulative distribution
functions of the LOS velocity $V_\parallel$ and of the
apparent spatial off-center displacement $L_{\bot}$
in host galaxies for the 15 objects in our sample
{ which have numerical solutions fulfilling observational constraints.
Here we give detailed comments for the objects one by one.}

\begin{description}
{

\item{J0001-0033}: a radio galaxy at z=0.247 (Figure~\ref{fig6}). 
The probability of having Doppler-shifted broad emission lines with $V_\parallel>200~\kms$  is $<1.5\%$, consistent with $0$ given the
uncertainties in our model assumptions.
{ The probability of having AGN off-center displacement 
with $\theta_\bot>0.1$ mas is about $40\%$.}
 The
critical velocity $V_{\rm dp}$ ranges from $450$ to $1300~\kms$, when
the uncertainties in $f_{\rm bol}$ and $f_{\rm tor}$ are considered. The
corresponding probability of being ``dust-poor'' ranges from $76\%$ to 
$12\%$. The critical velocity $V_{\rm cr}$ ranges from $2800$ to 
$4000~\kms$, resulting in a negligible probability of being ``dust-free''.

\item{J0049+0059}: a radio galaxy at z=0.304 (Figure~\ref{fig6}). 
 The probabilities of detecting Doppler-shifted broad emission lines
 with $V_\parallel>200~\kms$ and AGN off-center displacement with
 $\theta_\bot>10$ mas are consistent with 0. The probability of having
 AGN off-center displacement with $\theta_\bot>0.1$ mas is about
 $40\sim50\%$. The 
critical velocity $V_{\rm dp}$ ranges from $560$ to $1600~\kms$, when
the uncertainties in $f_{\rm bol}$ and $f_{\rm tor}$ are considered. The
corresponding probability of being ``dust-poor'' ranges from $64\%$ to 
$(1\sim2)\%$. The critical velocity $V_{\rm cr}$ ranges from $3600$ to 
$5100~\kms$, resulting in a negligible probability of being ``dust-free''.

\item{J0808+2409 (3C192)}: a radio galaxy at z=0.060
    (Figure~\ref{fig6}). 
    Notice that the distribution functions of $V_\parallel$ and $\theta_\bot$ 
    vary with $w_1$ by as large as an order of magnitude.
    The probability of having Doppler-shifted broad 
    lines with $V_\parallel>200~\kms$ is non-zero,
    but $<5\%$.
    The probability of having AGN
    off-center displacement with $\theta_\bot>0.1$ mas is about 50\%.
     The critical velocity $V_{\rm dp}$ ranges from 
    $240$ to $690~\kms$, when the uncertainties in $f_{\rm bol}$ 
    and $f_{\rm tor}$ are considered. The corresponding probability 
    of being ``dust-poor'' ranges from approximately $56\%$ to $89\%$.
    The critical velocity $V_{\rm cr}$ ranges from $1500$ to 
    $2200~\kms$, resulting in a negligible probability of being ``dust-free''.
    
\item{J0831+3219 (4C+32.25)}: a radio galaxy at z=0.051
    (Figure~\ref{fig6}). For $60^{\circ}
    < w_1 \leq 90^{\circ}$, none of the $10^8$ trial Monte Carlo 
    simulations produces a solution fulfilling all the observational 
    constraints. While for other $w_1$, the probabilities 
    of detecting Doppler-shifted broad emission lines with 
    $V_\parallel>200~\kms$ and AGN off-center displacement 
    with $\theta_\bot>10$ mas are consistent with 0. 
    The probability of having AGN
    off-center displacement with $\theta_\bot>0.1$ mas is about 50$\sim$63\%.
    The critical velocity 
    $V_{\rm dp}$ ranges from 
    $200$ to $550~\kms$, when the uncertainties in $f_{\rm bol}$ 
    and $f_{\rm tor}$ are considered. The corresponding probability 
    of being ``dust-poor'' ranges from approximately $66\%$ to $87\%$.
    The critical velocity $V_{\rm cr}$ ranges from $2000$ to 
    $5500~\kms$, resulting in a negligible probability of being ``dust-free''.

\item{J0859-0433}: a radio galaxy at z=0.356 (Figure~\ref{fig6}). 
    The probability of $V_\parallel>200~\kms$ is $<8\%$.
     The probabilities of have AGN
    off-center displacement with $\theta_\bot>10$ mas is $<6\%$,
    while the probability for $\theta_\bot>0.1$ mas ranges from $46\%$
    to $60\%$.
    We do not have data about the bolometric luminosity for this 
    object thus cannot give an estimation to the probability of
    being ``dust-poor'' or ``dust-free''. 

\item{J0924+4233}: a radio galaxy at z=0.227 (Figure~\ref{fig6}). 
    For $60^{\circ}
    < w_1 \leq 90^{\circ}$, none of the $10^8$ trial Monte Carlo 
    simulations gives a solution fulfilling all the observational 
    constraints. While for other $w_1$, the probability 
    of having Doppler-shifted broad emission lines with 
    $V_\parallel>200~\kms$ is consistent with 0, given the uncertainties
    in our model assumptions. The probability of having
    AGN off-center displacement with $\theta_\bot>10$ mas is
    non-zero, but $<8\%$, while the probability for $\theta_\bot>0.1$ mas
    ranges from 41\% to 54\%. The critical velocity 
    $V_{\rm dp}$ ranges from 
    $350$ to $980~\kms$, when the uncertainties in $f_{\rm bol}$ 
    and $f_{\rm tor}$ are considered. The corresponding probability 
    of being ``dust-poor'' ranges from approximately $30\%$ to $83\%$.
    The critical velocity $V_{\rm cr}$ ranges from $2200$ to 
    $3100~\kms$, resulting in a negligible probability of being ``dust-free''.
    
\item{J0941-0143}: a radio galaxy at z=0.384 (Figure~\ref{fig7}). 
    The probability of having Doppler-shifted broad emission lines with 
    $V_\parallel>200~\kms$ is $<15\%$. The probabilities of having AGN
    off-center displacement with $\theta_\bot>10$ mas is about 30\%
    and the probability for $\theta_\bot>0.1$ mas is about $57\%$.
    The critical velocity $V_{\rm dp}$ ranges from 
    $190$ to $550~\kms$, when the uncertainties in $f_{\rm bol}$ 
    and $f_{\rm tor}$ are considered. The corresponding probability 
    of being ``dust-poor'' ranges from approximately $67\%$ to $91\%$.
    The critical velocity $V_{\rm cr}$ ranges from $1200$ to 
    $1700~\kms$, and the resulting probability of being ``dust-free'' is
    $0\sim17\%$.

\item{J0941+3944 (3C223.1)}: a radio galaxy at z=0.107
    (Figure~\ref{fig7}). For $60^{\circ}
    < w_1 \leq 90^{\circ}$, none of the $10^8$ trial Monte Carlo 
    simulations gives a solution fulfilling all the observational 
    constraints. While for other $w_1$, the probabilities of having 
    Doppler-shifted broad 
    lines with $V_\parallel>200~\kms$ and AGN
    off-center displacement with $\theta_\bot>10$ mas are both
    consistent with $<8\%$, given the uncertainties in our
    model assumptions. 
    For $\theta_\bot>0.1$ mas, the probability is about $24\%$. 
    The critical velocity $V_{\rm dp}$ ranges from 
    $200$ to $580~\kms$, when the uncertainties in $f_{\rm bol}$ 
    and $f_{\rm tor}$ are considered. The corresponding probability 
    of being ``dust-poor'' ranges from $54\%$ to $89\%$.
    The critical velocity $V_{\rm cr}$ ranges from $2000$ to 
    $5800~\kms$. The resulting probability of being ``dust-free'' is
    $(0.01\sim4.4)\%$ for $20^\circ<w_1<40^{\circ}$ and
    $(0.08\sim12)\%$ for $40^\circ<w_1<60^{\circ}$.

\item{J1005+1154}: a radio galaxy at z=0.166 (Figure~\ref{fig7}).
    For $60^{\circ}
    < w_1 \leq 90^{\circ}$, none of the $10^8$ trial Monte Carlo 
    simulations produces a solution satisfying all the observational 
    constraints. While for other $w_1$, the probabilities 
    of showing Doppler-shifted broad emission lines with 
    $V_\parallel>200~\kms$ and AGN off-center displacement 
    with $\theta_\bot>10$ mas are consistent with 0, given the uncertainties
    in our model assumptions. 
    For $\theta_\bot>0.1$ mas, the probability ranges from $23\%$ to
    $42\%$. The critical velocity $V_{\rm dp}$ ranges from 
    $350$ to $1000~\kms$, when the uncertainties in $f_{\rm bol}$ 
    and $f_{\rm tor}$ are considered. The corresponding probability 
    of being ``dust-poor'' ranges from $25\%$ to $76\%$. 
    The critical velocity $V_{\rm cr}$ ranges from $2200$ to 
    $3200~\kms$, resulting in a negligible probability of 
    being ``dust-free''.
    
\item{J1020+4831 (4C+48.29)}: a radio galaxy at z=0.053
    (Figure~\ref{fig7}). The probability of detecting Doppler-shifted 
    broad emission lines with 
    $V_\parallel>200~\kms$ is consistent with 0, given the 
    uncertainties in our model assumptions.
    The probability of detecting AGN off-center displacement
    with $\theta_\bot>10$ mas is
    non-zero, but $<7\%$. For $\theta_\bot>0.1$ mas, the probability is about
    45\%. The critical velocity 
    $V_{\rm dp}$ ranges from 
    $220$ to $620~\kms$, when the uncertainties in $f_{\rm bol}$ 
    and $f_{\rm tor}$ are considered. The corresponding probability 
    of being ``dust-poor'' ranges from $58\%$ to $87\%$. 
    The critical velocity $V_{\rm cr}$ ranges from $1400$ to 
    $2200~\kms$. The resulting probability of being ``dust-free'' is
    $(0\sim11)\%$ for $20^\circ<w_1<40^{\circ}$,
    $(0\sim7.9)\%$ for $40^\circ<w_1<60^{\circ}$,
    and $(0\sim8.2)\%$ for $60^\circ<w_1<90^{\circ}$.

\item{J1130+0058 (4C+01.30)}: a radio quasar at z=0.132 (Figure~\ref{fig7}).
    For $60^{\circ}
    < w_1 \leq 90^{\circ}$, none of the $10^8$ trial Monte Carlo 
    simulations produces a solution satisfying all the observational 
    constraints. While for other $w_1$, the probability of
    $V_\parallel>500~\kms$ is $<11\%$, but the probability of
    $V_\parallel>200~\kms$ ranges from $15\%$ to $20\%$.
    The probability of detecting AGN off-center displacement is
    non-zero, ranging from $(16\sim24)\%$ in the case
    $20^\circ<w_1<40^{\circ}$ to $(25\sim33)\%$ in the case
    $40^\circ<w_1<60^{\circ}$.
    The critical velocity $V_{\rm dp}$ ranges from 
    $700$ to $990~\kms$, when the uncertainties in $f_{\rm bol}$ 
    and $f_{\rm tor}$ are considered. The corresponding probability 
    of being ``dust-poor'' ranges from $80\%$ to $96\%$. 
    The critical velocity $V_{\rm cr}$ ranges from $1100$ to 
    $3100~\kms$. The resulting probability of being ``dust-free'' is
    $(25\sim50)\%$ for $20^\circ<w_1<40^{\circ}$ and
    $(34\sim59)\%$ for $40^\circ<w_1<60^{\circ}$.
    It is interesting to note that \citet{wan03} reported a possible large 
    offset, $3.5~\arcsec$, in this object between 
    the radio core and the photometric center
    of the host galaxy. 
    Our Monte Carlo simulations suggest that 
    the observation would require a recoiling velocity
    of $ {V} >4000~\kms$ at a confidence level of $99.9\%$. Such high
    recoiling velocity is extremely difficult to achieve by asymmetric 
    gravitational wave radiation of circular SMBHB orbit. 
    This implies that either the two SMBHs merge along
    hyperbolic orbit \citep{hea09} or a mechanism other than
    recoiling SMBH is needed to produce the large displacement.
    
\item{J1140+1057}: a radio galaxy at z= 0.081 (Figure~\ref{fig8}). 
    The probability of having Doppler-shifted 
    broad emission lines with 
    $V_\parallel>200~\kms$ is generally $<8\%$.
    The probability of having AGN off-center displacement
    with $\theta_\bot>10$ mas is about $18\%$, increasing to about 
    $50\%$ for $\theta_\bot>0.1$ mas. The critical velocity 
    $V_{\rm dp}$ ranges from 
    $250$ to $700~\kms$, when the uncertainties in $f_{\rm bol}$ 
    and $f_{\rm tor}$ are considered. The corresponding probability 
    of being ``dust-poor'' ranges from $55\%$ to $87\%$. 
    The critical velocity $V_{\rm cr}$ ranges from $1600$ to 
    $2200~\kms$, resulting in a negligible probability of being
    ``dust-free''.
    
\item{J1207+3352}: a radio galaxy at z=0.079 (Figure~\ref{fig8}). 
    For $60^{\circ}
    < w_1 \leq 90^{\circ}$, none of the $10^8$ trial Monte Carlo 
    simulations produces a solution satisfying all the observational 
    constraints. While for other $w_1$, the probabilities of 
    showing Doppler-shifted broad emission lines with $V_\parallel>200~\kms$
    and AGN off-center displacement with $\theta_\bot>10$ mas are both
    $\le15\%$. For $\theta_\bot>0.1$ mas, the probability is
    about 43\%. The critical velocity 
    $V_{\rm dp}$ ranges from 
    $160$ to $440~\kms$, when the uncertainties in $f_{\rm bol}$ 
    and $f_{\rm tor}$ are considered. The corresponding probability 
    of being ``dust-poor'' ranges from approximately $67\%$ to $92\%$,
    The critical velocity $V_{\rm cr}$ 
    ranges from $980$ to $1400~\kms$. The resulting 
    probability of being ``dust-free'' is
    $(2.7\sim17)\%$ for $20^\circ<w_1<40^{\circ}$ and
    $(8.6\sim32)\%$ for $40^\circ<w_1<60^{\circ}$.

\item{J1348+4411}: a radio galaxy at z=0.267 (Figure~\ref{fig8}). 
     The probability of
    $V_\parallel>500~\kms$ is about $13\%$, but the probability of
    $V_\parallel>200~\kms$ increases steeply to about $35\%$.
    The probability of having AGN off-center displacement is
    ranging from about $52\%$ in the case of $\theta_\bot>10$ mas
    to about $63\%$ in the case of $\theta_\bot>0.1$ mas.
    The critical velocity $V_{\rm dp}$ ranges from 
    $89$ to $250~\kms$, when the uncertainties in $f_{\rm bol}$ 
    and $f_{\rm tor}$ are considered. The corresponding probability 
    of being ``dust-poor'' ranges from $85\%$ to $97\%$. 
    The critical velocity $V_{\rm cr}$ ranges from $560$ to 
    $800~\kms$. For all $w_1$, 
    the probability of being ``dust-free'' and become type I AGN is
     $(45\sim64)\%$.  
    The lack of broad line in this object implies that $V<V_{\rm cr}$. 
    If $V<V_{\rm cr}$ is required,
    our Monte Carlo simulations suggests that $P(V_{\parallel}>200~\kms)$
    and $P(\theta_{\bot}>0.1~{\rm mas})$ both decrease to $<10\%$,
    but $P_{\rm dp}$ is still in the range $(70\sim90)\%$. Alternatively, 
    it is possible that 
    this radio galaxy intrinsically lacks broad emission lines like in some
    AGNs and quasars \citep[e.g.][]{fan99}.

\item{J1614+2817}: a radio galaxy at z=0.108 (Figure~\ref{fig8}). 
    The probabilities of having Doppler-shifted broad emission lines 
    with $V_\parallel>200~\kms$ and AGN off-center displacement 
    with $\theta_\bot>10$ mas  are non-zero, but $<20\%$. For 
    $\theta_\bot>0.1$ mas, the probability increases to about $65\%$.
    The critical velocity $V_{\rm dp}$ ranges from 
    $610$ to $1700~\kms$, when the uncertainties in $f_{\rm bol}$ 
    and $f_{\rm tor}$ are considered. The corresponding probability 
    of being ``dust-poor'' ranges from nearly $0\%$ to $65\%$. 
    The critical velocity $V_{\rm cr}$ ranges from $3800$ to 
    $5400~\kms$, resulting in a negligible probability of being
    ``dust-free''.}
    
\end{description}

%%%%%%%%%%%%%%%%%%%%%
\section{Discussions and conclusions}
\label{sec:dc}

A popular model in the literature suggested that the X-shaped radio
sources form because of jet reorientation following the spin flip of
SMBHs at SMBHB mergers. Numerical relativity suggested that the black
hole merger not only leads to a spin flip but also to a gravitational
kick of the post-merger black holes. In this paper, we investigated with
Monte Carlo simulations the distribution function of the spin-flip
angles and recoiling velocities of the post-merger SMBHs, and the
detectability of recoiling SMBHs, in the forms of
Doppler-shifted spectral lines and off-center AGNs, 
in radio sources showing spin-flip signatures  {of jet reorientation}.
We also collected a sample of 19 X-shaped radio objects and
calculated for all the sample objects the spectral kinematic
offsets $V_{\parallel}$, the apparent spatial off-center displacements
$L_\bot$, and the degree of dust-torus truncation because of the
gravitational kick.

Because the spin flip and recoiling velocity strongly depend on the
parameters $q$, ${\bf a}_1$, ${\bf a}_2$, and $\bf{e_s}$ of SMBHB
system, we first investigated the distribution of mass ratio $q$ in
spin-flip radio sources with given apparent jet reorientation angles,
assuming that the spin parameter $a$ of jet-ejecting SMBHs has $a\ge
0.9$. Our results indicate that to cause significant apparent jet
 {reorientation} the mass ratio cannot be smaller than a minimum value
$q_{\rm min} \sim 0.1$. The larger the apparent jet reorientation
angle is, the larger the minimum mass ratio will be. For spin-flip
radio sources with apparent jet reorientation angle
$15^{\circ}<\delta<40^{\circ}$, $q_{\rm min} \sim 0.05$ and the
most probable mass ratio is $q \simeq 0.3$, while for spin-flip
radio sources with $65^{\circ} < \delta < 90^{\circ}$, $q_{\rm min}
\simeq 0.15$ and the most probable mass ratio is $q \ga 0.6$. Our
results suggest that if  {an observed jet reorientation} is indeed
due to the spin flip following a SMBHB merger, the merger must be
major merger with $q\ga 0.2-0.3$. If the spin-flip model for radio
sources  {of jet reorientation} 
could be confirmed, our results suggest that we can measure the
distribution of SMBHB mass ratio in luminous radio sources and give
statistic constraints on the hierarchical galaxy formation model and the
growth history of SMBHs.

 {Having} the mass ratio distribution, we investigated the 
distributions of spin flip angle and recoiling velocity 
in spin-flip radio sources. Our results showed that the larger
the SMBHB mass ratio is, the larger the the most possible spin flip
angle and the recoiling velocity will be. For typical major merger
mass ratio
$q=0.3$, $q=0.7$ and $q=1.0$, the most possible spin flip angles are
moderate and increase with mass ratio from $\Delta = 19^{\circ}$ for
$q=0.3$, and $25^{\circ}$ for $q=0.7$ to $30^{\circ}$ for $q=1$. The
spin flip angle is hardly ever larger than $\Delta \sim 55^{\circ}$. Our
results imply that any apparent jet reorientation angle $\delta \ga
60^{\circ}$ is due to projection effect. For $q = 0.3$, the recoiling
velocity ranges from about zero to larger than $1200\, {\rm
km~s^{-1}}$ with a most possible velocity $V \sim  600\, {\rm km
  s^{-1}}$. While for $q \ga 0.7$, the distribution of recoiling
velocity is similar and ranges from about $0$ to larger than $2100\,
{\rm km~s^{-1}}$ with most probable velocity $\hat{V}\sim 900\, {\rm
km~s^{-1}}$. For major mergers with $q\ga 0.3$, almost all recoiling
SMBHs move nearly along the final jet orientation within a angle of
$40^\circ$. Alignment  {of} spin axes with orbital angular momentum
of SMBHB prior to  {the black hole} coalescence would significantly
reduce the recoiling velocity and spin-flip angle, but tighten the
correlation between the orientations of final spin and recoiling
velocity. Our results
suggest that to efficiently detect recoiling SMHBs with Doppler
shifted spectral broad emission lines, one should preferentially
observe  {nearly face-on AGNs, e.g.} radio loud
quasars or blazars among which the broad emission lines 
are nearly maximumly Doppler-shifted for a given recoiling velocity. 
This conclusion is also
consistent with present observations in the literature that all
candidates for recoiling SMBHs are bright quasars and the measured
Doppler-shifting recoiling velocity is preferentially very larger
\citep[e.g.][]{kom08}. However, to detect recoiling SMBHs in radio
galaxies, one should try to measure
the apparent spatial off-center displacements of AGNs relative to the
centers of the host galaxies because the recoiling SMBHs are moving
nearly along the jet axes, which are at large angles with respect to
LOS. Even in some quasars, recoiling SMBHs may be discovered by
measuring the apparent spatial off-center displacements because some
radio galaxies and Type II quasars may  {transit} to Type I quasars
due to the recoiling oscillations as suggested by \citet{KM08a}. For
minor mergers with
$q\sim 0.1$, both the spin-flip angle and the recoiling velocity
narrowly distribute with peaks at the most possible angle $\Delta \sim
12^\circ$ and velocity $V\sim 150\, {\rm km~s^{-1}}$.

In order to use the observed jet reorientation angles in 
spin-flip radio sources to constrain the observable spectral
kinematic offsets $V_\parallel$ and apparent spatial off-center
displacements $L_\bot$, we projected the spin-flip angles and the
recoiling velocities along and vertical to LOS and investigated their
correlations. An analytical 1-D model is used to calculate the
deceleration of a recoiling SMBH due to dynamical friction, and the
elapsed time since the black hole gets kicked is estimated from the
size of the radio morphology. In general, $V_\parallel$ in spin-flip
radio sources is smaller than $2000~\kms$ and $L_\bot$  {hardly ever}
exceeds 2 kpc. However, we should note that the conclusion is valid
only for quasars with SMBHBs of circular orbits. For two SMBHs merging
along hyperbolic orbits, it is possible to have a recoiling velocity
significantly larger than $2000~\kms$ \citep{hea09}. Our results
indicate that in radio sources showing spin-flip signatures  {of
jet reorientation}, detectable Doppler-shifted broad emission lines
are mainly contributed by GPS and CSS sources with large apparent
spin-flip angles. The detection rate of Doppler-shifted broad emission
lines in a sample of GPSs/CSSs with $\delta>40^\circ$ could be as high
$50\%$. For GRSs, deceleration due to dynamical friction has
significantly reduced $V_\parallel$, so the probability of detecting
Doppler-shifted broad emission lines is typically $\la10\%$. 
Detecting AGN off-center displacement requires astrometric
observations with high angular resolution.  {For current telescopes
with limit resolution up to} $10$ mas, detection of off-center
displacement would be limited to the GPS/CSS sources within a 
distance of 100 Mpc.  {However,} future interferometer missions
with sub-mas resolution  {or higher}, such as Gaia, would allow
detection of off-center AGNs  {in GPS, CSS, and GRSs at} a much
larger distance. Our results suggest that to
efficiently detect the recoiling SMBHs in spin-flip radio sources, one
should select a sample of {\rm GPS/CSS} sources with apparent jet reorientation
angles $\delta \ga 40^\circ$.

When the SMBH in an AGNs gets kicked, the outer part of the dust torus in
AGN unification model may become unbound and the dust torus is
truncated, leading to the transition of the AGN from dust rich to ``dust
poor'' or even to ``dust free''. When a FRII radio galaxy or
an obscured quasar becomes ``dust free'', it should transform to an
unobscured broad emission line quasar. We calculated the minimum
recoiling velocities for partial or full truncation of dust torus
and estimated their detectability in typical spin-flip radio sources
with different apparent jet reorientation angles $\delta$. Our
results suggest that about $11\%$ spin-flip radio galaxies with $\delta
\ga 40^\circ$ and about $1\%$ with $\delta \la 40^\circ$ will become
``dust free'' and broad emission line AGNs, if broad line regions
  are present. Our results imply that quasars with larger
bolometric luminosity and smaller black hole mass should have higher
probability to become ``dust free''. The velocity for dust torus to be
partially truncated is typically lower than the most-probable
recoiling velocity, therefore
most X-shaped radio sources should be ``dust poor''. Quasars may
become weak emission line objects, if the broad emission
regions are partially truncated due to gravitational
recoiling. This is possible for a quasar with small black hole mass   
($\sim10^6~\msun$)
radiating at the Eddington limit. We note that without gas replenishing
the accretion disks, the dust-free and weak-line quasars formed by recoiling
SMBHs are relatively short-living ($\la1$ Myr), because the combination of
high recoiling velocity ($\ga10^3~\kms$), small black hole mass
($\la10^7~\msun$), and large bolometric luminosity ($\ga10^{45}~{\rm
  erg~s^{-1}}$) will significantly reduce the lifetimes of the
accretion disks around the recoiling black holes
\citep[e.g.][]{loe07}. This argument implies that among X-shaped radio
sources whose host  galaxies are typically gas-poor elliptical, the
ones with long active jets ($\ga 100$ kpc or $t_a \ga 3\times
10^6$ yr for a typical advancing speed $0.1c$ of hot spot in FRII radio 
galaxies) and small black hole masses ($\la 10^7~\msun$) are unlikely 
to host dust-free and weak-line quasars. 

We then selected a sample of 19 X-shaped radio objects, consisting of
18 X-shaped radio galaxies and 1 quasar, and for each sample
object we calculated the distribution functions of the LOS recoiling velocity
$V_\parallel$ and the apparent spatial off-center displacement $L_\bot$,
as well as the critical recoiling velocities $V_{\rm cr}$
and $V_{\rm dp}$ for partial and complete truncation of dust torus.
To do the calculations, we measured the apparent jet reorientation
angles from the radio images and collected the central SMBH masses,
the optical luminosities of AGNs, the viewing angles and dynamic ages
of the active lobes from the literature. Our results showed that most
X-shaped radio objects in the sample have negligible $V_\parallel$
($P_{v,\parallel}<10\%$), except for J0941-0143, 4C+01.30, J1348+4411,
and J1614+2817. All the 15 sample objects with valid Monte Carlo
solutions have significant probabilities ($>20\%$) to show AGN
off-center displacements when a spatial resolution of $0.1$ mas is
considered. In particular, the radio galaxy J1348+4411 have
appreciable probability to show both Doppler-shifted broad emission
lines ($P_{v,\parallel}\simeq40\%$) and off-center spatial
displacement ($P_{\theta,\bot}\simeq55\%$), therefore would be
intriguing target for follow-up observations. Our results also showed
that the sample X-shaped radio objects have high probabilities to be
``dust-poor''. The probabilities for X-shaped radio galaxy J1348+4411
and quasar 4C+01.30 are always higher than $80\%$, despite the
uncertainties in torus size and bolometric correction factor. Because
of the high probability of being both dust-poor and offset from galaxy
center, these two objects would be interesting  targets for IR
follow-up observations to test the dust-poorness. Besides, J1348+4411
and 4C+01.30 also have the highest probability in our sample to be
``dust-free'' and non-obscured. The fact that 4C+01.30 is already a
quasar implies that this object could be an obscured AGN before binary
black hole merger, but may have transformed into a quasar after recoil
takes place. Detecting the predicted off-center displacement and
Doppler shifted broad emission lines would strongly support that
4C+01.30 is a recoiling SMBH. For J1348+4411, the lack of broad line
in this object  suggests that the recoiling velocity is toward the
lower end of the predicted distribution function, but the probability
of being dust-poor is still higher than $70\%$. 

The detection of the predicted observational signatures of recoiling
SMBHs in X-shaped radio sources would give final confirmation of the
spin-flip model for X-shaped radio sources. While, if none of the
predicted signatures is detected in X-shaped radio sources, the SMBHB
spin-flip model for the formation of X-shaped radio feature would be
most likely excluded, unless our assumptions for the initial
conditions of SMBHB mergers are severely unrealistic. However, this
does not imply that X-shaped radio sources do not harbor SMBHBs at
center, because jet reorientation may form due to the interaction of
SMBHB and accretion disk \citep{liu04}. According to the disk-binary
interaction model, the final recoiling velocity of a SMBH is small due
to the alignment between the black hole spin axes and the orbital
angular momentum before the merger, therefore neither the spectral
kinematic offset nor the spatial AGN off-center displacement are as
large as those in the spin flip model.

\acknowledgments

{ We are grateful to the anonymous referee for valuable comments
which  {help us to} significantly improve the quality of this paper.}
We also thank Luciano Rezzolla, Joan Centrella, Stefanie Komossa,
Xuebing Wu, Monica Colpi, Zhiqiang Shen 
for helpful discussions. This work is
supported by the Chinese national 973 program (2007CB815405), the
National Natural Science Foundation of China (NSFC11073002), and the
China Scholarship Council for financial support (2009601137). DW
thanks the support of the Maoyugang Undergraduate Research
Fellowship. The numerical computations were carried out on the SGI Altix $330$
system at the Astronomy Department, Peking University.

\clearpage
\begin{figure}
\plotone{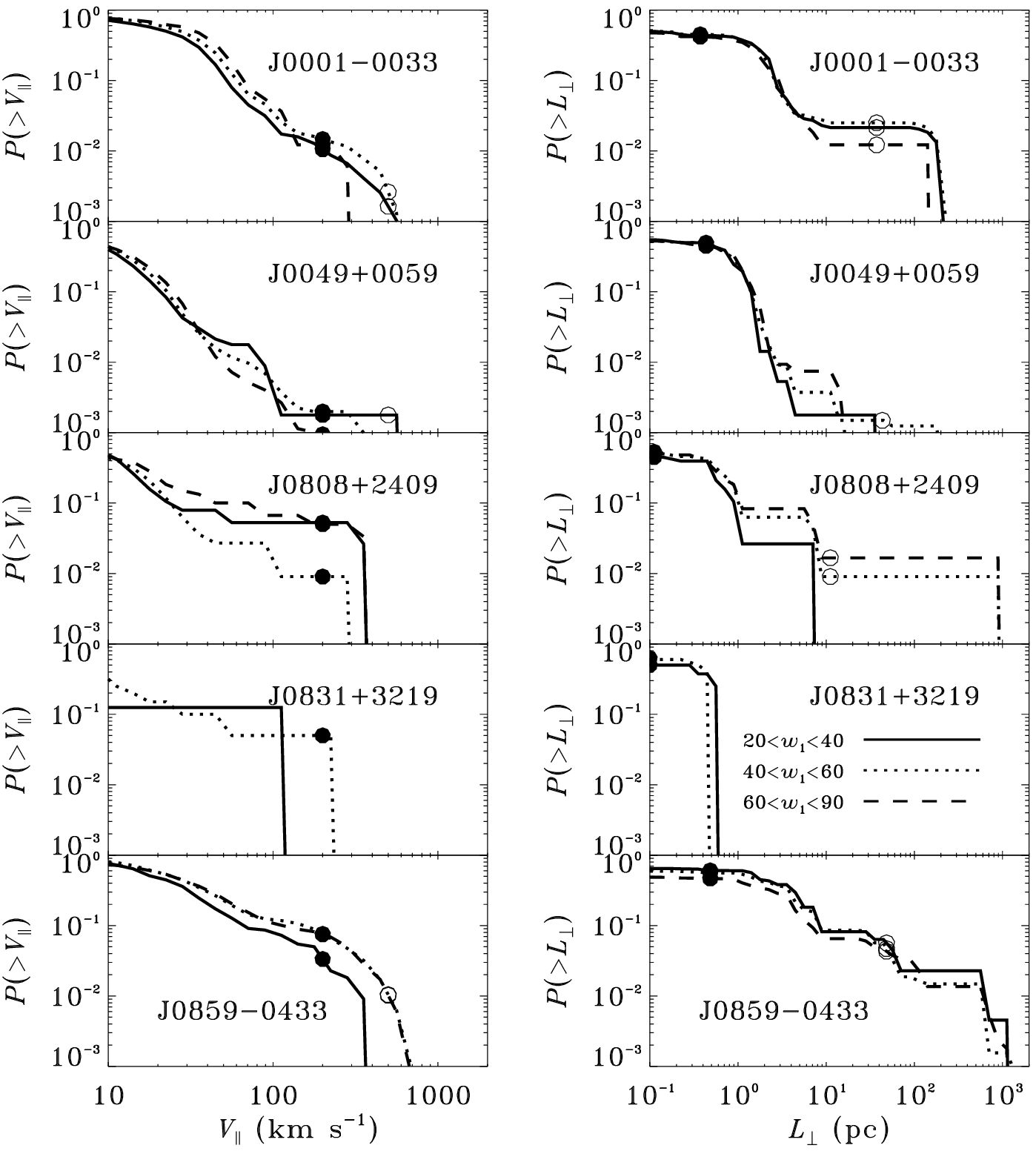}
\caption{Cumulative distribution of LOS velocity
  $V_{\parallel}$ ({\it Left}) and apparent spatial off-center
  displacements $L_{\bot}$ vertical to LOS ({\it right}) for X-shaped
  radio objects J0001-0033, J0049+0059, 3C192, J0831+3219, 
  and J0859-0433. The dots and circles in the left panels mark, respectively,
   the locations of $V_\parallel=200~\kms$ and $500~\kms$, while those
   in the right panels mark, respectively, the locations of $\theta_\bot=10$
   mas and $0.1$ mas.
\label{fig6}}
\end{figure}
\clearpage

\begin{figure}
\plotone{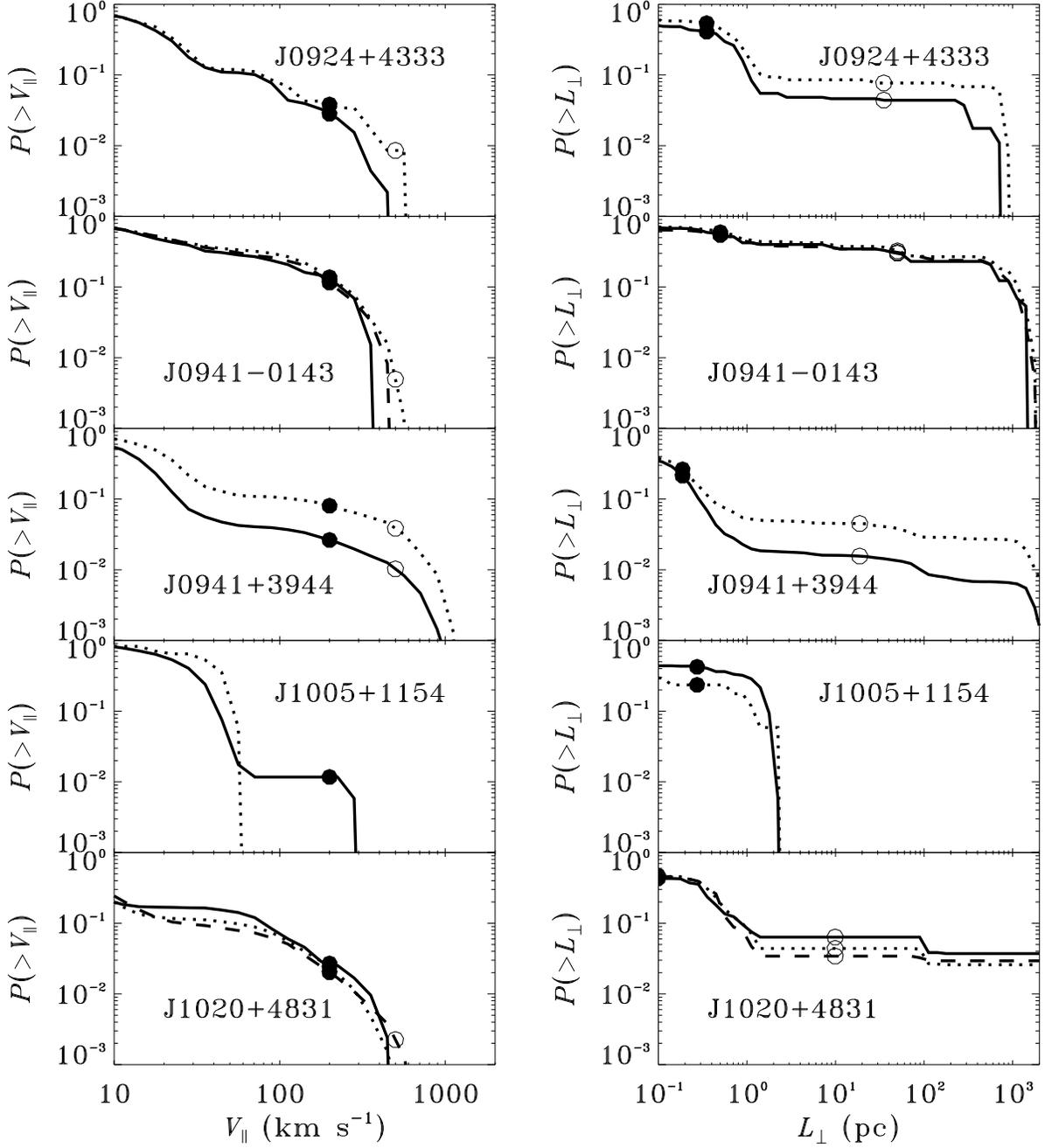}
\caption{The same as Figure~\ref{fig6} but for
  objects J0924+4333, J0941-0143, 3C223.1, J1005+1154, 
  4C+48.29.
\label{fig7}}
\end{figure}
\clearpage

\begin{figure}
\plotone{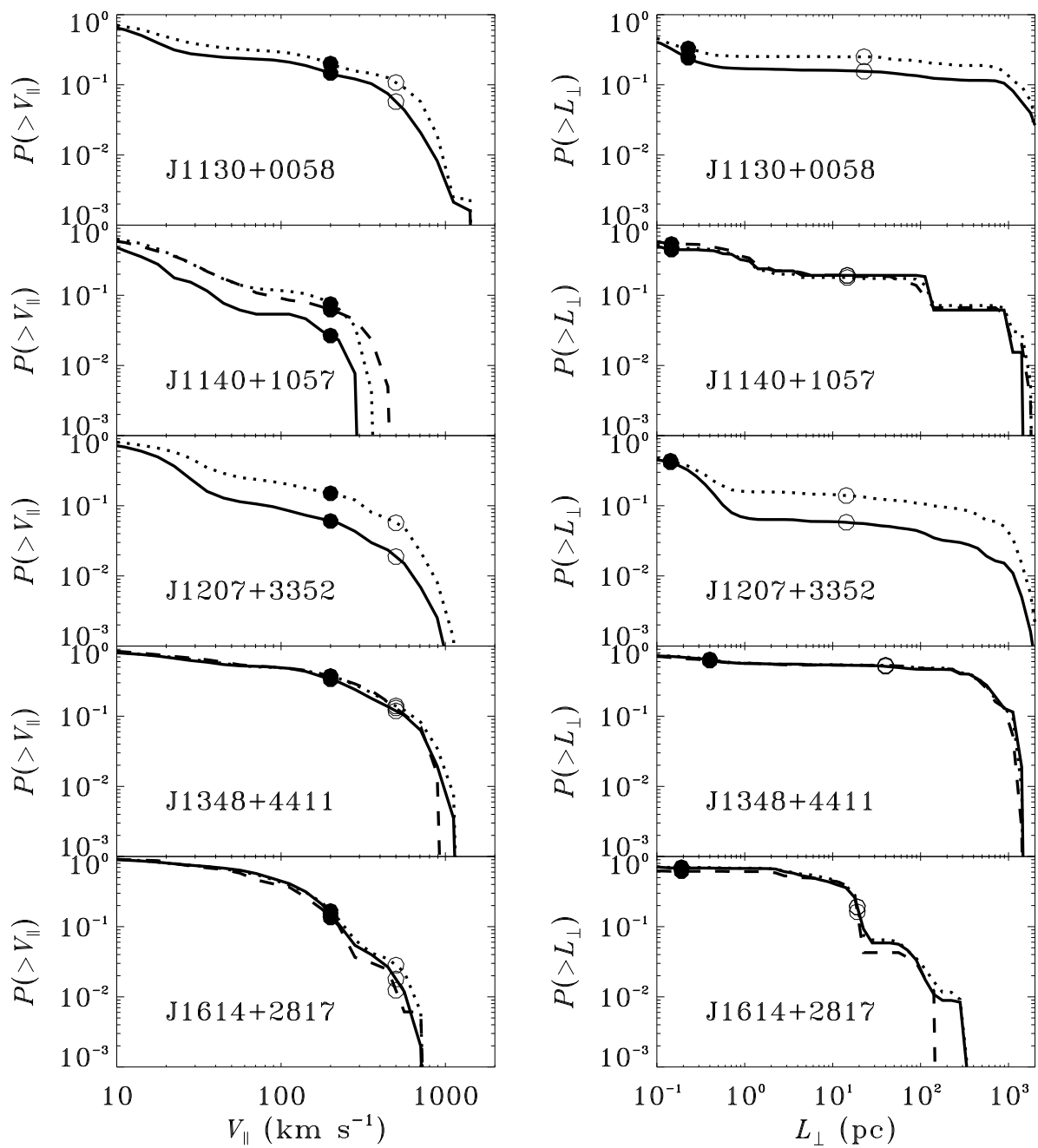}
\caption{The same as Figure~\ref{fig6} but for objects
  4C+01.30, J1140+1057, J1207+3352, J1348+4411, 
  and J1614+2817.
\label{fig8}}
\end{figure}
\clearpage

\begin{deluxetable}{cccccccccccccc}
\tabletypesize{\scriptsize}%{\footnotesize}
\tablecolumns{15}
\tablewidth{0pt}
\rotate
\tablecaption{X-shaped samples\label{tab:sample}}
\tablehead{
\colhead{IAU Name} &
\colhead{ID z} &
\colhead{$\log\,M$} &
\colhead{$\log\, \lambda L_\lambda(5100{\rm\AA})$} &
\colhead{$\delta$} &
\colhead{$w_2$} &
\colhead{$\log{t_a}$} &
\colhead{ref} &
\colhead{$V_{\rm {cr}}$} &
\colhead{$V_{\rm {dp}}$} &
\colhead{$w_1$} &
\colhead{$P_{v,\parallel}$} &
\colhead{$P_{\theta,\bot}$}      &
\colhead{$P_{\rm dp}$} \\
        & & \colhead{[$M_\odot$]} & \colhead{[$\rm{erg~s^{-1}}$]} & \colhead{[$^{\circ}$]} & \colhead{[$^{\circ}$]} & \colhead{[yr]} &  &  \colhead{[$10^2~\rm{km~s}^{-1}$]} &  \colhead{[$10^2~\rm{km~s}^{-1}$]} & \colhead{[$^{\circ}$]} & [\%] & [\%] & [\%] \\
(1) & (2) & (3) & (4) & (5) & (6) & (7) & (8) & (9) & (10) & (11) & (12) &(13) & (14)
}
\startdata
J0001$-$0033 &G 0.247 & $8.70$ & 43.77 & $60\pm7.5$ & 45$\sim$70 & 6.37 & 1,3,4 & $34$ & $7.5$               & 20$\sim$40 & $0.16, 1.1$ & $2.1, 43$ & $50_{13}^{72}$\\
   & &  &  &  &  & & & & & 40$\sim$60 & $0.26, 1.5$ & $2.5, 45$ & $50_{14}^{71}$ \\
   & &  &  &  &  &  & & & & 60$\sim$90 & $0.0, 1.2$ & $1.2, 42$  & $55_{12}^{76}$ \\
   \\
J0049$+$0059 &G 0.304 & $8.73$ & 43.43 & $60\pm20$ & 60$\sim$90 & 6.68 & 1,3,4 & $42$ & $9.5$ & 20$\sim$40 & $0.18,0.19$  & $0.0,50$ & $33_{2.0}^{64}$ \\
   & &  &  &  &  &  & & & & 40$\sim$60 & $0.0,0.20$ & $0.15,49$ & $34_{1.6}^{64}$ \\
   & &  &  &  &  &  & & & & 60$\sim$90 & $0.031,0.10$ & $0.015,45$ & $33_{1.7}^{64}$ \\
   \\
J0808$+$2409 &G 0.060 & $8.21$ & 43.85 & $60\pm7.5$ & 60$\sim$90 & 6.52 & 2,3,4 & $18$ & $4.1$                  & 20$\sim$40 & $0.0,5.3$ & $0.0,45$  & $77_{57}^{89}$ \\
3C192  &  &  &  &  &  &  & & & & 40$\sim$60 & $0.0,0.90$ & $0.90,53$ & $80_{56}^{88}$ \\
   & &  &  &  &  &  & & &             & 60$\sim$90 & $0.0,5.0$ & $1.7,50$ & $80_{61}^{88}$ \\
   \\
J0831$+$3219 &G 0.051 & $8.08$ & 43.97 & $65\pm7.5$ & 60$\sim$90 & 6.69 & 2,3,4 & $15$ & $3.3$                      &20$\sim$40 & $0.0,0.0$ & $0.0,50$ &  $87_{76}^{87}$ \\
4C+32.25 & &  &  &  &  &  & & & & 40$\sim$60 & $0.0,5.0$ & $0.0,63$ & $80_{66}^{85}$   \\
  & &  &  &  &  &  & & &                  & 60$\sim$90 & \nul & \nul & \nul \\
  \\
J0859$-$0433 &G 0.356 & $8.52$ & \nod & $60\pm15$ & 60$\sim$90 & 6.08 & 1,3,4 & \nod & \nod                     & 20$\sim$40 & $0.0,3.3$ & $5.7,60$ & \nod\\
  & &  &  &  &  &  & & & & 40$\sim$60 & $1.0,7.6$ & $4.7,56$ &  \nod\\
  & &  &  &  &  &  & & & & 60$\sim$90 & $1.0,7.4$ & $4.346$ &  \nod\\
  \\
J0924$+$4233 &G 0.227 & $8.38$ & 43.58 & $70\pm7.5$ & 45$\sim$70 & 6.54 & 1,3,4 & $26$  & $5.8$    & 20$\sim$40 & $0.0,2.8$ & $4.4,41$ & $63_{30}^{79}$  \\
  & &  &  &  &  &  & & & & 40$\sim$60 & $0.85,3.8$ & $7.7,54$ &  $64_{33}^{83}$ \\
  & &  &  &  &  &  & & & & 60$\sim$90 & \nul & \nul & \nul   \\
  \\
J0941$-$0143 &G 0.384 & $7.68$ & 43.19 & $55\pm7.5$ & 60$\sim$90 & 6.14 & 1,3,4 & $15$ & $3.3$     & 20$\sim$40 & $0.0,13$ & $30,55$  &  $84_{68}^{90}$ \\
  & &  &  &  &  &  & & & & 40$\sim$60 & $0.49,14$ & $32,59$ &  $82_{67}^{91}$  \\
  & &  &  &  &  &  & & & & 60$\sim$90 & $0.0,12$ & $30,55$ &  $82_{67}^{90}$ \\
  \\
J0941$+$3944 &G 0.107 & $8.10$ & 43.94 & $65\pm7.5$ & 15$\sim$35 & 6.44 & 2,3,4 & $15$ & $3.4$                    & 20$\sim$40 & $1.0,2.6$ & $1.5,21$ &  $75_{54}^{86}$ \\
 3C223.1 & &  &  &  &  &  & & & & 40$\sim$60 & $3.9,8.0$ & $4.5,26$ & $80_{62}^{89}$  \\
  & &  &  &  &  &  & & &                 & 60$\sim$90 & \nul & \nul & \nul \\
  \\
J1005$+$1154 &G 0.166 & $8.67$ & 44.12 & $75\pm7.5$ & 45$\sim$70 & 6.43 & 1,3,4 & $26$ & $6.0$     & 20$\sim$40 & $0.0,1.1$ & $0.0,42$ & $58_{25}^{76}$ \\
  & &  &  &  &  &  & & & & 40$\sim$60 & $0.0,0.0$ & $0.0,23$ & $51_{32}^{59}$  \\
  & &  &  &  &  &  & & & & 60$\sim$90 & \nul  & \nul & \nul \\
  \\
J1020$+$4831 &G 0.053 & $8.08$ & 43.78 & $45\pm7.5$ & 60$\sim$90 & 6.72 & 2,3,4 & $16$ & $3.7$                      & 20$\sim$40 & $0.0,2.7$ & $6.3,43$ &  $76_{59}^{85}$ \\
4C+48.29 &  &  &  &  &  &  & & & & 40$\sim$60 & $0.050,2.4$ & $4.3,46$ &  $77_{58}^{87}$ \\
  & &  &  &  &  &  & & &                   & 60$\sim$90 & $0.22,2.0$ & $3.4,47$  &  $78_{60}^{87}$ \\
  \\
J1101$+$1640 &G 0.069 & $8.30$ & 43.69 & $72\pm7.5$ & 60$\sim$90 & 6.62 & 2,3,4 & $22$ & $5.0$ & 20$\sim$40 & \nul & \nul & \nul \\
  & &  &  &  &  &  & & & & 40$\sim$60 & \nul & \nul & \nul \\
  & &  &  &  &  &  & & & & 60$\sim$90 & \nul & \nul & \nul \\
  \\
J1130$+$0058& Q 0.132 & $7.65$ & 44.10 & $80\pm7.5$ & 15$\sim$35 & 6.24 & 3,4,5 & $8.3$ & $1.9$                   & 20$\sim$40 & $5.8,15$ & $16,24$ & $89_{80}^{95}$ \\
4C+01.30 & &  &  &  &  &  & & & & 40$\sim$60 & $11,20$ & $25,33$ & $89_{81}^{96}$\\
  & &  &   &  &  &  & & &                & 60$\sim$90 & \nul & \nul & \nul \\
  \\
J1140$+$1057 & G 0.081 & $8.10$ & 43.61 & $55\pm7.5$ & 60$\sim$90 & 6.28 & 1,3,4 & $18$ & $4.1$     & 20$\sim$40 & $0.0,2.7$ & $19,45$ & $78_{57}^{87}$ \\
  & &  &  &  &  &  & & & & 40$\sim$60 & $0.0,7.5$ & $18,47$ &  $76_{57}^{87}$ \\
  & &  &  &  &  &  & & & & 60$\sim$90 & $0.0,6.2$ & $19,54$  &  $76_{55}^{87}$\\
  \\
J1207$+$3352 &G 0.079 & $7.96$ & 44.13 & $65\pm7.5$ & 15$\sim$35 & 6.19 & 1,3,4 & $12$ & $2.6$    & 20$\sim$40 & $1.9,6.0$ & $5.8,41$ & $82_{67}^{91}$ \\
  & &  &  &  &  &  & & & & 40$\sim$60 & $5.7,15$ & $14,44$ &  $85_{74}^{92}$  \\
  & &  &  &  &  &  & & & & 60$\sim$90 & \nul & \nul & \nul \\
  \\
J1210$-$0341 &G 0.178 & $8.30$ & 43.90 & $82\pm7.5$ & 45$\sim$70 & 5.97 & 1,3,4 & $20$ & $4.4$ & 20$\sim$40 & \nul & \nul & \nul \\
  & &  &  &  &  &  & & & & 40$\sim$60 & \nul & \nul & \nul \\
  & &  &  &  &  &  & & & & 60$\sim$90 & \nul & \nul & \nul \\
  \\
J1327$-$0203 &G 0.183 & $8.43$ & 44.01 & $80\pm7.5$ & 60$\sim$90 & 6.39 & 1,3,4 & $22$ & $4.8$ & 20$\sim$40 & \nul & \nul & \nul\\
  & &  &  &  &  &  & & & & 40$\sim$60 & \nul & \nul & \nul\\
  & &  &  &  &  &  & & & & 60$\sim$90 & \nul & \nul & \nul \\
\\  
J1348$+$4411 &G 0.267 & $7.05$ & 43.28 & $65\pm15$ & 45$\sim$70 & 5.87 & 1,3,4 & $6.7$ & $1.5$ & 20$\sim$40 & $12,33$ & $51,63$ & $93_{86}^{97}$ \\
  & &  &  &  &  &  & & & & 40$\sim$60 & $14,37$ & $53,64$ & $91_{85}^{96}$ \\
  & &  &  &  &  &  & & & & 60$\sim$90 & $13,37$ & $53,62$ & $94_{88}^{97}$ \\
\\
J1444$+$4147 &G 0.188 & $8.34$ & 43.74 & $68\pm7.5$ & 60$\sim$90 & 6.60 & 1,3,4 & $23$ & $5.1$ & 20$\sim$40 & \nul & \nul  & \nul \\
  & &  &  &  &  &  & & & &40$\sim$60 & \nul & \nul & \nul \\
  & &  &  &  &  &  & & & & 60$\sim$90 & \nul & \nul & \nul \\
\\  
J1614$+$2817 & G 0.108 & $9.08$ & 44.02 & $60\pm7.5$ & 45$\sim$70 & 5.59 & 1,3,4 & $45$ & $10$     & 20$\sim$40 & $1.8,15$ & $19,69$ &  $29_{0.78}^{61}$ \\
  & &  &  &  &  &  & & & & 40$\sim$60 & $2.8,17$ & $19,70$ &  $29_{0.92}^{60}$ \\
  & &  &  &  &  &  & & & & 60$\sim$90 & $1.2,13$ & $16,62$ & $33_{0.0}^{65}$ \\
\enddata
\tablenotetext{a}{there are no solutions in our Monte Carlo calculations with $10^8$ random number experiments}
\tablenotetext{b}{This source belongs to X-shaped quasar}
\tablenotetext{c}{no SDSS photometry available}
\tablerefs{
(1) \citet{che07}, (2) \citet{lal07}, (3) \citet{lan10}, (4)
  \citet{mez11}, (5) \citet{wan03}}
\end{deluxetable}

\appendix

\section{Deceleration of recoiling SMBH due to dynamical friction}\label{decel}

{ We set up an 1-D analytic model to calculate the deceleration of
  a recoiling SMBH due to dynamical friction against stars and dark
  matter (DM).  We adopted a core-S\'{e}rsic law \citep{ter05} for
  the stellar density distribution and an NFW profile for DM halo
  \citep{vol03}. The normalization of the density is determined by the
  conditions that (1) the mass ratio between SMBH and host galaxy is
  $M_\bullet/M_g=0.002$ and (2) the mass deficit \citep{mer06a} inside
  core radius equals the mass of SMBH.  {Interested} readers are
  referred to \citet{gua08} for  {more detailed discussions on} the
  density distribution of host galaxy. 

 {For a recoiling spuermassive black hole with mass $M_\bullet$
at galactic centers $r=0$ obtaining an initial kick velocity $v=v_k$, 
the N-body numerical simulations \citep{mer06a,li11} suggests that the
dynamical evolution of recoiling black hole within the host galaxy and
dark halo can be divided into three phases and is well described by
the equations incorporating} the Chandrasekhar formula for dynamical
friction 
\begin{eqnarray}
  {dr\over dt}&=&v\\
  \frac{dv}{dt}&=&-\frac{GM_g(r)}{r^2}-4\pi\ln\Lambda
  G^2M_\bullet\rho(r)\frac{v}{|v|^3}\left[{\rm
      erf}(X)-\frac{2X}{\sqrt{\pi}}e^{-X^2}\right], 
\end{eqnarray}
where $M_g(r)$ is the total mass inside the sphere of radius $r$ about
the galaxy center, $\ln\Lambda$ is the Coulomb logarithm, $\rho(r)$ is
the total density of stars and dark matter, $X=v/\sqrt{2}\sigma(r)$,
and $\sigma(r)$ is the stellar velocity dispersion at $r$.  {The
N-body simulation given by \citet{mer06a} shows that when the total
mass within the orbit of the oscillating black hole is larger than the
black hole mass $M_\bullet$,  the dynamic evolution of recoiling black
hole is at Phase I and the} fiducial Coulomb logarithm is 
$\ln\Lambda=2.5$. However, when the stellar mass within the orbit of
black hole become smaller than $M_\bullet$,  {the dynamic evolution
of recoiling black hole goes into an long-term oscillatory Phase II
during which the Coulomb logarithm is in the range $0.1 \la \ln\Lambda
\la 0.3$ with a typical value $\ln\Lambda \simeq 0.2$ \citep[Figure
  10 of][]{gua08}}. At a much later time of Phase III, the black hole
reaches a thermal equilibrium with stars and has a Brownian motion at
the galaxy center \citep{gua08}. For the cases which we are
interested in, the black hole never goes into Phase III and we will
not considered it here.

Figure~\ref{df} shows the resulting evolution of 
$r$ as a function of time $t$, where $r$ is normalized by the
effective radius of the host galaxy, $R_e$, and $t$ is normalized by
the dynamical timescale $T_e$ at $R_e$. Different trajectories are for
different initial kick velocity $v_k$, which is normalized by the
escape velocity $v_{\rm esc}$ of the host galaxy and listed in the
legend. Our model is essentially the same as Model A2 in
\citet{gua08}, except that the ratio $M_\bullet/M_g$ in our model is
two times greater, which results in a slightly longer dynamical
friction timescale than that in \citet{gua08}. 

\begin{figure}
\plotone{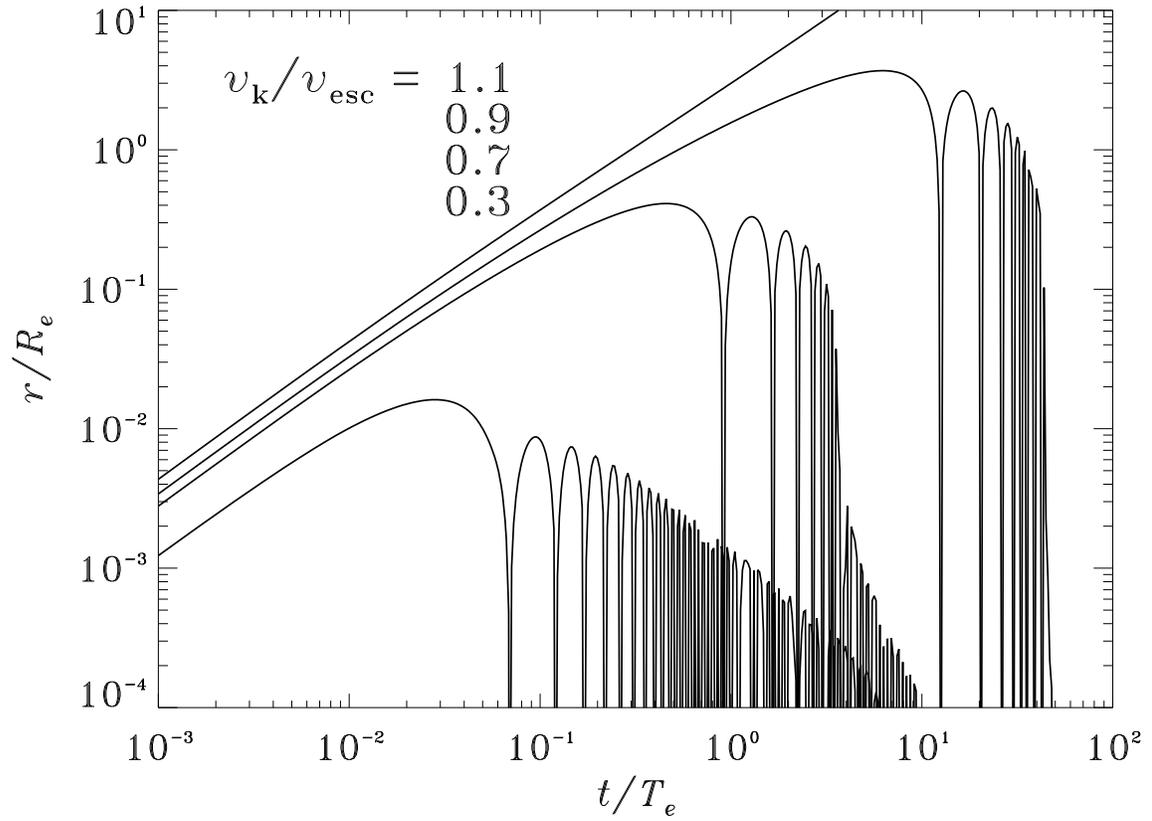}
\caption{Trajectories of recoiling SMBH with different initial kick
  velocity. Because of the logarithmic scale, the oscillatory phase
    II \citep{gua08} is clearly resolved in the plot when $v_k/v_{\rm
      esc}=0.3$ and $0.7$.
\label{df}} 
\end{figure}

Notice that in our  {simple} analytic model, the {\it normalized}
trajectories of  {$r/R_e$($t/T_e$) and $v/v_{\rm esc}$($t/T_e$)}
are uniquely determined by $v_k/v_{\rm ese}$ and do not depend on
$M_\bullet$. This characteristics 
allows us to derive the trajectories of $r(t)$ and $v(t)$ for any kick
velocity $v_k$ and black hole mass $M_\bullet$, by interpolating among
a limited number of {\it normalized} trajectories pre-calculated for
different $v_k/v_{\rm esc}$ values. The normalized trajectories that
we used in our Monte Carlo simulations are for $v_k/v_{\rm esc}=0.01,\
0.05, \ 0.1,\ 0.2,\ 0.3,\ ...,\ 2.0$. 
}

\end{document}